\documentclass[apj,onecolumn,numberedappendix]{emulateapj}

\usepackage{epsfig,amsmath} 

\eqsecnum

\newcommand{\eqpair}{{\sc eqpair}} 

\newcommand{\beq}{\begin{eqnarray}}
\newcommand{\eeq}{\end{eqnarray}}
\newcommand{\bez}{\begin{eqnarray*}}
\newcommand{\eez}{\end{eqnarray*}}
\newcommand{\bc}{\begin{center}}
\newcommand{\ec}{\end{center}}

\newcommand{\vnb}{\mbox{\boldmath $\nabla$}}
\newcommand{\vomega}{\mbox{\boldmath $\omega$}}
\newcommand{\vOmega}{\mbox{\boldmath $\Omega$}}

\newcommand{\rmd}{{d}}

\newcommand{\unb}{\underline{\nabla}}

\newcommand{\UB}{U_{\rm B}}
\newcommand{\alphafs}{\alpha_{\rm f}}
\newcommand{\Bcr}{B_{\rm cr}}
\newcommand{\tstar}{\tau_*}
\newcommand{\taua}{\tau_{\rm a}}
\newcommand{\taus}{\tau_{\rm sc}}
\newcommand{\alphaa}{\alpha_{\rm a}}
\newcommand{\alps}{\alpha_{\rm sc}}

\newcommand{\vecx}{\mbox{\boldmath $x$}}
\newcommand{\vecp}{\mbox{\boldmath $p$}}
\newcommand{\vecpp}{\mbox{\boldmath $p$}_{+}}
\newcommand{\vecpm}{\mbox{\boldmath $p$}_{-}} 
\newcommand{\fourx}{\underline{x}}
\newcommand{\fourp}{\underline{p}}
\newcommand{\fourpp}{\underline{p}_{\,+}}
\newcommand{\fourpm}{\underline{p}_{\,-}}

\newcommand{\dotn}{\dot{n}} 
\newcommand{\dotN}{\dot{N}}

\newcommand{\dotgammas}{\dot{\gamma}_{\rm s}} 
\newcommand{\dotgammac}{\dot{\gamma}_{\rm c}} 
\newcommand{\dotxc}{\dot{x}_{\rm c}}

\newcommand{\Nphtwo}{{N_{\rm ph}^2}}
\newcommand{\Nph}{{N_{\rm ph}}}
\newcommand{\nph}{{n_{\rm ph}}}

 \newcommand{\nocc}{\tilde{n}} 
 \newcommand{\noccx}{\tilde{n}_{\rm ph}} 
  \newcommand{\noccpr}{\tilde{n}_{\rm ph}}
  \newcommand{\tesc}{t_{\rm esc}}   
  
  \newcommand{\Rph}{R_{\rm ph}}
    \newcommand{\Sph}{\overline{R}_{\rm ph}}
   \newcommand{\overRph}{\overline{R}_{\rm ph}}
  \newcommand{\Aph}{A_{\rm ph}}
    \newcommand{\Bph}{B_{\rm ph}}
 \newcommand{\Kph}{K_{\rm ph}}
    \newcommand{\Dph}{D_{\rm ph}}

  \newcommand{\gridx}{\Delta_x}
   \newcommand{\deltax}{\delta_x} 
 \newcommand{\gridp}{\Delta_p}
    
  \newcommand{\distr}{n}
  
  \newcommand{\me}{m_{\rm e}}
    \newcommand{\lambdac}{\lambda_{\rm C}}
 
    \newcommand{\noccm}{\tilde{n}_{-}}
  \newcommand{\noccp}{\tilde{n}_{+}} 
 \newcommand{\nocce}{\tilde{n}_{\rm e}}
 	 \newcommand{\noccpm}{\tilde{n}_{\pm}}

 \newcommand{\nel}{{n_{\rm e}}}
 \newcommand{\nelmi}{{n_{\rm -}}}
  \newcommand{\npos}{n_{\rm +}}
 
	 \newcommand{\npoel}{{n_{\pm}}}
   
\newcommand{\Nel}{{N_{\rm e}}}
\newcommand{\Nelmi}{{N_{\rm -}}}
\newcommand{\Npos}{N_{\rm +}}
\newcommand{\Nplmn}{N_{\pm}}
\newcommand{\Nmnpl}{N_{\mp}}

 \newcommand{\Rel}{R_{\rm e}}
\newcommand{\overRe}{\overline{R}_{\rm e}}
 \newcommand{\Se}{\overline{R}_{\rm e}} 
  \newcommand{\Ae}{A_{\rm e}}
  \newcommand{\Be}{B_{\rm e}}
    \newcommand{\Ke}{K_{\rm e}}
       \newcommand{\De}{D_{\rm e}} 
       \newcommand{\Dacc}{D_{\rm acc}}

\newcommand{\Opl}{\Omega_{+}}

  \newcommand{\alphas}{\alpha_{\rm s}}
  
  \newcommand{\epssyn}{\epsilon_{\rm s}}

\newcommand{\gpl}{\gamma_{+}}
\newcommand{\gmn}{\gamma_{-}}

\newcommand{\bpl}{\beta_{+}}
\newcommand{\bmn}{\beta_{-}}

\newcommand{\zpl}{p_{+}}
\newcommand{\zmn}{p_{-}}
\newcommand{\zplmn}{p_{\pm}}

\newcommand{\gcm}{\gamma_{\rm cm}}
\newcommand{\bcm}{\beta_{\rm cm}}
 \newcommand{\xcm}{x_{\rm cm}}

\newcommand{\re}{r_{\rm e}}
\newcommand{\sigmat}{\sigma_{\rm T}}
\newcommand{\sigmapp}{\sigma_{\rm pp}}
\newcommand{\sigmapa}{\sigma_{\rm pa}}
\newcommand{\Fpp}{F_{\gamma\gamma}}

   \newcommand{\Rgg}{R_{\rm \gamma\gamma}}
   \newcommand{\dotgammaCoul}{\dot{\gamma}_{\rm Coul}} 
   \newcommand{\DCoul}{D_{\rm Coul}} 
   \newcommand{\gammarel}{\gamma_{\rm rel}} 
   \newcommand{\prel}{p_{\rm rel}}

\newcommand{\vecpprime}{\mbox{\boldmath $p$}'}
\newcommand{\vecponeprime}{\mbox{\boldmath $p$}'_1}
\newcommand{\fourpprime}{\underline{p}'}
\newcommand{\fourponeprime}{\underline{p}'_{1}}
\newcommand{\Fcc}{F_{\rm Coul}}
   
\begin{document}

\title{Time-dependent modeling of radiative processes in hot magnetized plasmas}

\shorttitle{Radiative processes in hot magnetized plasma}
\shortauthors{Vurm \& Poutanen}

\author{Indrek Vurm\altaffilmark{1} and Juri Poutanen}

\affil{Astronomy Division, Department of Physical Sciences, P.O.Box 3000, 
90014 University of Oulu, Finland;  indrek.vurm@oulu.fi, juri.poutanen@oulu.fi }
 \altaffiltext{1}{Also at Tartu Observatory, 61602 T\~{o}ravere, Tartumaa, Estonia}


\begin{abstract}
Numerical simulations of radiative processes in magnetized compact sources such as hot accretion disks around black holes, relativistic jets in active galaxies and gamma-ray bursts are complicated because the particle and photon distributions span many orders of magnitude in energy, they also strongly depend on each other, the  radiative processes behave significantly differently depending on the energy regime, and finally due to the enormous difference in the time-scales of the processes. We have developed a novel computer code for the time-dependent simulations that overcomes these problems.  The processes taken into account are Compton scattering, electron-positron pair production and annihilation, Coulomb scattering as well as synchrotron emission and absorption. No approximation has been made on the corresponding rates.   For the first time, we solve coupled integro-differential kinetic equations for photons and electrons/positrons without any limitations on the photon   and lepton energies. A numerical scheme is proposed to guarantee energy conservation when dealing with  synchrotron processes in electron and photon equations. We apply the code to model non-thermal pair cascades in the blackbody radiation field, to study the synchrotron self-absorption as particle thermalization mechanism, and to simulate time evolution of stochastically heated pairs and corresponding synchrotron self-Compton photon spectra which might be responsible for the prompt emission of gamma-ray bursts. Good agreement with previous works is found in the parameter regimes where comparison is feasible, with the differences attributable to our improved treatment of the microphysics. 
\end{abstract}

\keywords{accretion, accretion disks --- galaxies: active --- gamma rays: bursts  ---
gamma rays: theory --- radiation mechanisms: non-thermal ---   X-rays: binaries}

%

\section{Introduction}

Spectral energy distributions of a number of compact, magnetized, high-energy sources such as relativistic jets from active galaxies, gamma-ray bursts, black hole accretion disk-coronae are strongly affected and shaped by Compton scattering, synchrotron radiation and electron-positron pair production \citep[see e.g.][]{G99,gcc02,ZG04,SP04}.  Understanding the physical conditions in these sources requires detailed modeling of the interactions between the particles and photons, which is not an easy task. The basic problem is that we cannot compute radiative processes from a given a priori lepton distribution (e.g. Maxwellian or a power-law), because it  depends strongly on the radiation field, which in its turn is determined by the particle distribution. Another problem is that the time-scales for various processes differ by orders of magnitude. The energy range of particles and photons responsible for the emission also spans many orders of magnitude, with different processes making dominant contributions to the emergent spectrum in different bands. One of the main difficulties in calculating radiative processes over a wide range of energies is that a particular radiative process may behave significantly differently depending on the energy regime, the most well-known example of such processes being also the most important in relativistic plasma, namely Compton scattering. Depending on the energies of the interacting particles, an electron or a photon can lose (or gain) a significant or negligible fraction of its initial energy in one scattering. The former case has to be accounted for by the integral scattering terms in the kinetic equations, while the latter necessitates the Fokker-Planck treatment.

The treatment of radiative processes in relativistic plasmas has been the subject of several works. There are two basic approaches: Monte Carlo methods \citep[e.g.][]{SBS95,PS97} and solving the relevant kinetic equations \citep[e.g.][]{LZ87,coppi92,coppi99,NM98,PW05}. Both have their own advantages and disadvantages. Monte Carlo treatment makes it easy to take into account radiative transfer effects, on the other hand it usually suffers from poor photon statistics at high energies. Another problem can arise at very low energies, where the optical thickness to synchrotron absorption can be enormous. In the kinetic theory approach photon statistics is not an issue, the sole difficulty lies in solving the relevant integro-differential equations. In this work we have chosen to follow the second approach.

Due to the difficulties in solving the exact Boltzmann equations of the kinetic theory, different simplifying approximations have been made in earlier works. They fall in three basic categories: ad hoc assumptions about the particle energy distributions, approximate treatment of different physical processes, and  simplified treatment of radiative transport. Various approximations invoked to simplify the treatment of radiative processes at the same time limit the range of their applicability. One commonly employed approximation concerns Compton scattering, which is assumed to take place in the Thomson regime \citep[e.g.][ hereafter GHS98]{GHS98} and is accounted for by a simple cooling term in the electron equation. This sets two restrictions that the photon energy in the electron rest frame is smaller than the electron rest mass
and 
the average photon energy is much lower than the electron kinetic energy. Otherwise all photons would not contribute to electron cooling, the higher energy ones being downscattered via Compton scattering. This means that one is unable to treat cases when Comptonization approaches saturation, which may be relevant at high compactnesses, and to study electron heating by external radiation. Other works account for Klein-Nishina corrections to the electron cooling rate, but still neglect the diffusive nature of the process when electron and photon energies are comparable \citep{coppi92,MSC05,PW05}, which works towards establishing an equilibrium Maxwellian distribution. Another useful approximation, when the integral terms describing Compton scattering are  accounted for, is to consider ultrarelativistic electrons 
and very low energy photons \citep[e.g.][]{ZDZ88,MSC05}.
This, however, becomes increasingly inaccurate when the electrons cool to sufficiently low energies.
 	
The cyclo-synchrotron process also exhibits qualitatively different behavior depending on the energy of the radiating particles. If the emitting particles are relativistic, the emission spectrum is smooth and can span several orders of magnitude in energy, while in the nonrelativistic case the energy is radiated at discrete cyclotron harmonics and most of this radiation might be strongly self-absorbed. In the first case, the radiating particle (electron or positron) mostly loses its energy in a continuous fashion, while in the second case it can gain energy by absorbing the  cyclo-synchrotron photons emitted by other particles. This process  is a dominant particle thermalization mechanism in compact magnetized sources \citep{GGS88}. Its proper account requires accurate emissivities in the transrelativistic regime, because electron thermalization usually takes place at mildly relativistic energies. Some codes for computing radiative processes in relativistic plasma \citep[e.g. {\sc eqpair} described in][]{coppi92,coppi99} neglect this process completely as the electrons are {\it assumed} to be thermal at low energies or account for thermalization   by Coulomb  collisions only \citep{NM98}. In other approaches synchrotron thermalization is computed (GHS98), but Compton scattering is then considered only approximately. 

Owing to the fact that proper treatment of transport processes for all types of particles would make the task prohibitively difficult, and partly due to our ignorance of the exact geometry of the problem, it is rather common practice to neglect radiative transport altogether \citep[e.g.][]{LZ87,coppi92} and assume spatially homogeneous and isotropic particle distributions. In this case particle and photon loss from the system is modeled in terms of simple escape probabilities. We also follow this approach here. 

In this paper we introduce and describe in details a numerical code that can deal with Compton scattering, synchrotron emission and absorption, electron-positron pair production and annihilation without limitations on the energies of the photons and electrons/positrons. We solve coupled  integro-differential kinetic equations describing time evolution of the photons and lepton distributions simultaneously. When necessary, the Fokker-Planck  differential terms are substituted instead of the  integral terms with coefficients computed exactly from the moments of the integral equation.  Particle thermalization by synchrotron self-absorption, Coulomb (M{\o}ller) scattering as well as Compton scattering is considered. Extreme caution is taken when dealing with  synchrotron self-absorption, because of cancellation of large, almost equal terms, which can result in inaccuracies and huge energy sinks. Numerical simulations show that our code conserves energy with about 1\% accuracy. We present an extensive testing of the code using some problems described previously in the literature. Processes involving bremsstrahlung can be easily added to the code, while for the conditions considered in the paper, they are not important. 

\section{The kinetic equations}

We are considering a region of relativistic plasma of charged particles (electrons and positrons, which we call ``electrons'' below if the relevant processes, e.g.   Compton scattering and synchrotron, operate identically on both types of particles) permeated by radiation and tangled magnetic fields. We study the evolution of lepton and photon distributions by solving the time-dependent coupled kinetic equations accounting for synchrotron emission and absorption, Compton scattering, Coulomb scattering, and electron-positron pair production and annihilation. We make a simplifying assumption of homogeneity and isotropy of the particle distributions. We assume that energy is transferred to electrons by some unspecified mechanism, which is manifested as either injection of high-energy electrons or diffusive acceleration within the active region. The escape of radiation (and also electrons) from the region is modeled by a simple escape probability formalism.

\subsection{Distribution functions} 

The dimensionless four-momentum of a photon  is $\fourx =\{ x, \vecx \}= x \{ 1,\vomega\}$, where $\vomega$ is the unit vector in the photon propagation direction and $x \equiv h\nu/\me c^2$. The photon distribution  can be described by the occupation number $\noccx$ or by the photon number density per linear and logarithmic interval of photon energy: 
\begin{equation} 
\Nph = \int \Nph(x) \; \rmd x = \int \nph(x) \; \rmd\ln{x} = 
\frac{2}{\lambdac^3} \int \rmd^2 \omega \int \noccx(\vecx)\ x^2\ \rmd x ,
\end{equation}
where  $\lambdac=h/\me c$ is the Compton wavelength. Functions $\Nph(x)$ and $\noccx$ are used in general forms of kinetic equations and $\nph(x)$ is convenient for numerical work. 

The dimensionless electron (positron) four-momentum is defined  as $\fourp  = \{ \gamma, \vecp\}= \{ \gamma, p\vOmega\}= \gamma \{ 1, \beta\vOmega\} $, where $\vOmega$ is the unit vector in the electron propagation direction, $\gamma$, $\beta$, and $p=\beta\gamma=\sqrt{\gamma^2-1}$  are the electron Lorentz factor, dimensionless velocity and momentum, respectively. We can use subscripts $+$ and $-$ to distinguish between positrons and electrons. The electron/positron distributions can be defined in a number of alternative ways (normalized to their number density): 
\begin{equation} 
\Nplmn  =\int \Nplmn (\gamma) \rmd\gamma = \int \npoel(p) \; \rmd\ln{p} =  \frac{2}{\lambdac^3}  \int \rmd^2 \Omega \int \noccpm(\vecp) \; p^2 \rmd p. 
\end{equation}
The occupation number $\noccpm(p)$
and the density per unit Lorentz factor are useful quantities 
used in general kinetic equations, while the electron density per logarithmic momentum interval $\npoel(p)$  is more appropriate for numerical work.  
For the processes, where the distinction between electrons and positrons is unnecessary, we use the sum of the distributions, for example, $\nel = \nelmi + \npos$.
%

\subsection{General form of the kinetic equations} 

The relativistic kinetic equation (RKE) describing the evolution of the occupation number $\nocc_1 (\vecp_1)$ of species 1 (electron  or photon) as a result of binary collisions can be written in the covariant form \citep{dGvLvW80} 
\begin{equation}\label{eq:rke}
\fourp_1 \cdot \unb \nocc_1(\vecp_1)= 
\int \frac{\rmd^3 p_2}{\epsilon_2}\frac{\rmd^3 p_3}{\epsilon_3} \frac{\rmd^3 p_4}{\epsilon_4} 
 \delta(\fourp_1+\fourp_2-\fourp_3-\fourp_4) W_{12\rightarrow 34}
\left[ \nocc_3(\vecp_3)  \nocc_4(\vecp_4)-  \nocc_1(\vecp_1)  \nocc_2(\vecp_2) \right] ,
\end{equation}
where $\unb=\{\partial/c \partial t, \vnb\}$ is the four-gradient, $\epsilon_i$ is the zeroth component 
of the corresponding four-momentum,
and $W_{12\rightarrow 34}=W_{34\rightarrow 12}$ 
is a Lorentz scalar transition rate, which possesses the obvious symmetry. 
In this equation, the non-linear terms related to fermion degeneracy and induced photon scattering are omitted. 
As it stands, the right-hand side of equation~(\ref{eq:rke}) accounts for the rate of one particular process.
To determine the full evolution of $\nocc_1$ we should therefore sum up the collisional integrals accounting for all relevant processes.

In the frame where the particle distributions are isotropic (we call this frame $E$), 
the kinetic equation can be represented in the form (skipping subscript 1): 
\begin{equation}\label{eq1}
\frac{\partial \nocc(p)}{\partial t} + \frac{1}{p^2}\frac{\partial}{\partial p}\left\{
\, \dot{\epsilon} \, \epsilon \ p \, \nocc(p) - \frac{1}{2} \frac{\partial}{\partial \epsilon} \left[ D(\epsilon) \, \epsilon \ p \, \nocc(p) \right] \right\}
= \left. \frac{D \nocc(p)}{Dt} \right|_{\rm coll},			
\end{equation}
where the momentum derivative term accounts for continuous energy gain/loss processes,
while the right hand side contains all discontinuous processes such as scattering, emission, absorption and escape.
The quantities $\dot{\epsilon}$ and $D(\epsilon)$ 
account for systematic particle heating/cooling and diffusion in energy space, respectively.
Both  
are generally energy-dependent for the processes we are considering here.
%
For the following discussion it is convenient to decompose the kinetic equations in terms of the contributions from
different physical processes as
\beq \label{eq:kinph}
\frac{\partial \nph(x)}{\partial t} & =&  \dotn_{\rm ph,syn} (x)+ \dotn_{\rm ph,cs}(x)
+ \dotn_{\rm ph,pp} (x)- \frac{\nph(x)}{\tesc} + Q_{\rm ph}, \\ 
  \label{eq:kinel}
\frac{\partial \npoel(p)}{\partial t} & =& \dotn_{\rm \pm,syn}(p) +\dotn_{\rm \pm,cs}(p) + \dotn_{\rm \pm,pp} (p)  
+ \dotn_{\rm \pm,Coul} (p) - \frac{\npoel(p)}{t_{\rm \pm, esc}} + Q_{\rm \pm}, 
\eeq 
where syn, cs, pp and Coul stand for synchrotron,  Compton scattering, pair production (and annihilation), and Coulomb scattering, respectively.  The terms describing physical processes can contain both differential and integral parts, depending on the nature of the process and the way we find most convenient to treat it.
Thus the equation for photons has the form: 
\begin{equation} \label{eq:rke_gen_ph}
\frac{\partial \nph(x)}{\partial t} = -\frac{\partial}{\partial \ln x} \left[ \Aph(x) \nph(x) - \Bph(x) \frac{\partial \nph(x)}{\partial \ln x} \right] 
+ \int  \Kph(x,x_1) \nph(x_1) \: \rmd \ln x_1 - \frac{\nph(x)}{t_{\rm ph}}  +  S_{\rm ph}  .
\end{equation}
Here the differential term is responsible for Compton scattering in diffusion approximation, while 
the integral term with kernel $\Kph$ describes scattering that can be resolved on the grid. 
The sink term $\propto 1/t_{\rm ph}$  describes photon absorption (by synchrotron and pair-production) and scattering as well as the escape, while $S_{\rm ph}$ gives the contribution from pair annihilation, synchrotron emission and other (e.g. blackbody) photon  injections. 

Similarly for electrons and positrons, we write 
\begin{equation} \label{eq:rke_gen_el}
\frac{\partial \npoel(p)}{\partial t} = -\frac{\partial}{\partial \ln p} \left[ \Ae(p) \npoel(p) - \Be(p) \frac{\partial \npoel(p)}{\partial \ln p} \right] 
+ \int  \Ke(p,p_1) \npoel(p_1) \: \rmd \ln p_1 - \frac{\npoel(p)}{t_{\rm \pm}}  + S_{\rm \pm} , 
\end{equation}
where coefficients $\Ae$ and $\Be$ describe electron cooling, heating and diffusion as a result of synchrotron emission and absorption, Compton scattering in Thomson limit, Coulomb scattering as well as possible diffusive particle acceleration. 
The integral term with kernel $\Ke$ describes Compton scattering in Klein-Nishina limit into the bin and the sink term $\propto 1/t_{\pm}$ gives the scattering from the bin as well as the electron escape and pair annihilation. 
The source term  $S_{\pm}$ contains pair production as well as a possible electron injection term.

\subsection{Escape probability formalism}

As we are studying radiative processes in a simple one-zone framework neglecting the radiative transport effects,
we must include an escape term in equation (\ref{eq:kinph}) to allow for the fact that photons can leave the emission region of finite 
size  $R$ and produce the radiation flux that is actually observed. The typical escape timescale is usually estimated from random walk arguments resulting in $\tesc \sim R(1 + \taus)/c$, where $\taus$ is the scattering opacity. Such form accounts for the fact that if multiple scatterings are important ($\taus > 1$), photons have to 'diffuse' out of the medium and the escape time is prolonged by a factor $\taus$. However, it does not account for the fact that if the medium is absorptive, a typical photon cannot diffuse further than the thermalization length $l_{\star} = [\alphaa(\alphaa+\alps)]^{-1/2}$ before it is destroyed
($\alphaa$ and $\alps$ are extinction coefficients due to absorption and scattering, respectively).
To incorporate both effects, we employ the solution of a simple radiative diffusion problem
in a sphere of radius $R$ with constant emissivity, absorptivity and monochromatic scattering.
The escape timescale is estimated by comparing the emergent flux to the radiation density inside the source.
While clearly an oversimplification, such estimation nevertheless has the desired properties mentioned above.

Defining the effective optical thickness of the medium as $\tstar = \sqrt{3 \taua (\taua + \taus)}$, where $\taua= \alphaa  R$ and $\taus= \alps R$ are optical thicknesses due to absorption and scattering, respectively, we find
\begin{equation}\label{eq:tstar}
\tesc = \frac{2R}{3c} \left\{ 1 + \frac{\sqrt{3}}{2\sqrt{1-\lambda}}
\left[\frac{\tstar\left(1-{\rm e}^{-2\tstar}\right)}{\tstar \left( 1+{\rm e}^{-2\tstar} \right) - 
\left( 1-{\rm e}^{-2\tstar} \right)} - \frac{3}{\tstar}
\right]
\right\},														
\end{equation}
where $\lambda = \alps/(\alphaa + \alps)$ is the single-scattering albedo.
If the medium is translucent ($\tstar \ll 1$), equation (\ref{eq:tstar}) reduces to a more familiar form
\begin{equation} 
\tesc = \frac{2R}{3c} \left( 1 + \frac{3}{10} \taus \right).
\end{equation}
In our simulations, $\alphaa$ includes cyclo-synchrotron absorption and photon-photon pair production, and 
$\alps$ is the extinction coefficient for Compton scattering.  

\subsection{Compton scattering}

\subsubsection{Compton scattering of photons}

The explicitly covariant form  of RKE for Compton scattering of photons ignoring non-linear terms is 
(\citealt{Pom73}; \citealt{NP94}, hereafter NP94)  
\begin{equation} \label{eq:cs_ph}
\fourx \cdot \unb \noccx(\vecx)= \frac{\re^2}{2} \frac{2}{\lambdac^3}
\int \frac{\rmd^3 p}{\gamma} \frac{\rmd^3 p_1}{\gamma_1}  \frac{\rmd^3 x_1}{x_1} 
\: \delta(\fourp_1 + \fourx_1 - \fourp - \fourx)   \: F 
\left[ \noccx(\vecx_1)  \nocce(\vecp_1)-  \noccx(\vecx) \nocce(\vecp) \right] ,
\end{equation}
where $\re$ is the classical electron radius, $F$ is the Klein-Nishina reaction rate \citep{LLVol4}
\begin{equation}
F = \left( \frac{1}{\xi} - \frac{1}{\xi_1}\right)^2 + 2 \; \left( \frac{1}{\xi} - \frac{1}{\xi_1}\right)
+ \frac{\xi}{\xi_1} + \frac{\xi_1}{\xi},
\end{equation}
and $\xi = \fourp_1\cdot\fourx_1= \fourp\cdot\fourx$ and $\xi_1 =  \fourp_1\cdot\fourx =  \fourp\cdot\fourx_1$ are the scalar products of four-vectors.

We assume the existence of a reference frame where the particle and photon distributions are approximately homogeneous and isotropic.
Under the spacial homogeneity assumption we can write equation~(\ref{eq:cs_ph}) as
\begin{equation} \label{eq2}
\left. \frac{D\noccx(\vecx)}{Dt} \right|_{\rm coll,cs} =
- c \: \sigmat \: \overline{s}_0(\vecx) \: \Nel \: \noccx(\vecx)
+ c\sigmat \Nel \frac{1}{x} \int \frac{\rmd^3 x_1}{x_1} \: \Rph(\vecx_1 \rightarrow \vecx) \: \noccpr(\vecx_1).
\end{equation}
The scattering cross-section (in units of Thomson cross-section $\sigmat$) is given by
\begin{equation}\label{eq3}
\overline{s}_0(\vecx) = \frac{3}{16\pi} \frac{2}{\lambdac^3\Nel} \frac{1}{x}
\int \frac{\rmd^3 p}{\gamma} \frac{\rmd^3 p_1}{\gamma_1} \frac{\rmd^3 x_1}{x_1} \, \nocce(\vecp) \: F
\: \delta(\fourp_1 + \fourx_1 - \fourp - \fourx) 
\end{equation}
and the redistribution function is
\begin{equation} \label{eq4}
\Rph(\vecx_1 \rightarrow \vecx) = \frac{3}{16\pi} \frac{2}{\lambdac^3\Nel} 
\int \frac{\rmd^3 p}{\gamma} \frac{\rmd^3 p_1}{\gamma_1}  
\: \nocce(\vecp_1) \; F \; \delta(\fourp_1 + \fourx_1 - \fourp - \fourx) .  									
\end{equation}

For isotropic particle distributions in frame $E$, equation~(\ref{eq2}) can be written as
\begin{equation} \label{eq8}
\left. \frac{D \noccx(x)}{Dt} \right|_{\rm coll,cs} =
- c \: \sigmat \: \overline{s}_0(x) \: \Nel \: \noccx(x)
+ c\sigmat \Nel \frac{4\pi}{x} \int x_1 \rmd x_1 \: \Sph(x,x_1) \: \noccpr(x_1) ,
\end{equation}
where the redistribution function averaged over the   cosine of the scattering angle 
$\mu= \vecx \cdot\vecx_1/(x x_1) = \vomega \cdot \vomega_1$ is 
expressed via an integral over the electron distribution (NP94):
\begin{equation} \label{eq10}
\Sph(x,x_1) = \frac{1}{2} \int_{-1}^{1} \Rph(\vecx_1 \rightarrow \vecx) \: \rmd\mu 
= \frac{3}{16} \frac{2}{\lambdac^3\Nel} \:  \int_{\gamma_{\star}(x,x_1)}^{\infty} 
\overRph(x,x_1,\gamma_1) \, \nocce(p_1) \ \rmd \gamma_1 .	
\end{equation}
Here
\begin{equation}\label{eq12}
\overRph(x,x_1,\gamma_1) = 
\frac{1}{4\pi^2}\: p_1 \int \frac{\rmd^3 p}{\gamma} \rmd^2\Omega_1 \: \rmd^2\omega_1 \: F \:
\delta(\fourp _1 + \fourx_1 - \fourp  - \fourx) ,	
\end{equation}
and the lower limit of the second integral in equation~(\ref{eq10}) comes from the condition of energy and momentum conservation:
\begin{equation}\label{eq13}
\gamma_{\star}(x,x_1) = 
\begin{cases}
[ x - x_1 + (x+x_1)\sqrt{1+1/xx_1} ]/2   &\text{if $|x-x_1| \ge 2xx_1$,}   \\
1 + \left(x - x_1 + |x - x_1| \right)/2 &  \text{if $|x-x_1| \le 2xx_1$.}
\end{cases}						
\end{equation}
The integrals in equation~(\ref{eq12}) can be calculated analytically (\citealt{Bri84}; NP94) to obtain a fully 
general expression for $\overRph(x,x_1,\gamma_1)$ valid in all regimes (see Appendix \ref{app:crf}).
This is an alternative form of the function derived by \citet{Jones68}.

Since the total number of particles is conserved in Compton scattering, multiplying the rhs of equation~(\ref{eq8}) by $x^2$ integrating over $\rmd x$ must give zero, implying a relation between the redistribution function and the extinction coefficient (NP94)
\begin{equation}\label{eq13a}
 \overline{s}_0(x) = \frac{4\pi}{x} \int_0^{\infty}  \Sph(x_1,x) \; x_1  \; \rmd x_1 \ .	
\end{equation}
This can also be inferred directly from the definitions (\ref{eq3}) and (\ref{eq4}).

In the kinetic equation (\ref{eq:kinph}) for the photon density $\nph(x)$ the Compton
term is obtained by multiplying equation (\ref{eq8}) by $8\pi\lambdac^{-3}  x^3$.

\subsubsection{Compton scattering of electrons and positrons}
\label{sec:cse}

The description of Compton scattering for electrons and positrons is very similar to that for photons. In the linear approximation the RKE reads
\begin{equation} \label{eq:cs_el}
\fourp \cdot \unb \noccpm(\vecp)= \frac{\re^2}{2} \frac{2}{\lambdac^3}
\int\frac{\rmd^3 x}{x}  \frac{\rmd^3 x_1}{x_1}  \frac{\rmd^3 p_1}{\gamma_1}
\: \delta(\fourp_1 + \fourx_1 - \fourp - \fourx)   \: F 
\left[ \noccx(\vecx_1)  \noccpm(\vecp_1)-  \noccx(\vecx) \noccpm(\vecp) \right] .
\end{equation}
Neglecting spatial gradients,  equation~(\ref{eq:cs_el}) becomes
\begin{equation} \label{eq14}
\left. \frac{D \noccpm(\vecp)}{Dt} \right|_{\rm coll,cs} = - c \: \sigmat \: \overline{s}_0(\vecp) \: \Nph \: \noccpm(\vecp)
+ c \sigmat \Nph \frac{1}{\gamma} \int \frac{\rmd^3 p_1}{\gamma_1} \: \Rel(\vecp_1 \rightarrow \vecp) \: \noccpm(\vecp_1),
\end{equation}
where the scattering cross-section for electrons is
\begin{equation}\label{eq15}
\overline{s}_0(\vecp) = \frac{3}{16\pi} \frac{2}{\lambdac^3\Nph} \frac{1}{\gamma} 
\int \frac{\rmd ^3 x}{x} \frac{\rmd ^3 x_1}{x_1} \frac{\rmd ^3 p_1}{\gamma_1} \noccx(\vecx) \: F
\: \delta(\fourp_1 + \fourx_1 - \fourp - \fourx)					 
\end{equation}
and the redistribution function
\begin{equation}  \label{eq16}
\Rel(\vecp_1 \rightarrow \vecp) = \frac{3}{16\pi} \frac{2}{\lambdac^3\Nph}
 \int \frac{\rmd ^3 x}{x} \frac{\rmd ^3 x_1}{x_1} 
\: \noccpr(\vecx_1) \; F  \; \delta(\fourp_1 + \fourx_1 - \fourp - \fourx) .	
\end{equation}
Making use of the isotropy of the problem, we can rewrite the kinetic equation in frame $E$ for isotropic 
distribution $\nocce(p)$: 
\begin{equation} \label{eq18}
\left. \frac{D \noccpm(p)}{Dt} \right|_{\rm coll,cs} = - c \: \sigmat \: \overline{s}_0(p) \: \Nph \: \noccpm(p)
+ c \sigmat \Nph \frac{4\pi}{\gamma} \int p_1 \rmd \gamma_1 \: \Se(p,p_1) \: \noccpm(p_1), 
\end{equation}
where the electron redistribution function averaged over cosine of the electron scattering angle
$\mu_e$ is
\begin{equation} \label{eq19}
\Se(p,p_1) = \frac{1}{2} \int_{-1}^{1} \Rel(\vecp_1 \rightarrow \vecp) \: \rmd\mu_e  = 
\frac{3}{16} \: \frac{2}{\lambdac^3\Nph} \int_{x_{\star}(\gamma,\gamma_1)}^{\infty} 
\overRe(\gamma,\gamma_1,x_1) \,
\noccpr(x_1) \ \rmd x_1 ,							
\end{equation}
where
\begin{equation} \label{eq20}
\overRe(\gamma,\gamma_1,x_1) = \frac{1}{4\pi^2} \: x_1 \int \frac{\rmd ^3 x}{x} \rmd^2\omega_1 
\: \rmd^2\Omega_1 \: F \: \delta(\fourp _1 + \fourx_1 - \fourp  - \fourx) 
\end{equation}
and
\begin{equation}\label{eq21}
x_{\star}(\gamma,\gamma_1) = [\gamma - \gamma_1 + |p - p_1|]/2.		
\end{equation}
The relation between the redistribution function and the extinction coefficient is
\begin{equation} \label{eq22a}
\overline{s}_0(p) = \frac{4\pi}{\gamma} \int  \Se(p_1,p) \, p_1 \, \rmd \gamma_1.
\end{equation}

Not surprisingly, there turns out to be a relation between the quantities $\overRph$ and $\overRe$ 
(proved in Appendix \ref{app:relation}), namely
\begin{equation}\label{eq22b}
pp_1 \overRe(\gamma,\gamma_1,x_1) = xx_1 \overRph(x,x_1,\gamma_1),			
\end{equation}
together with the energy conservation condition $x + \gamma = x_1 + \gamma_1$.
Through equations~(\ref{eq22b}) and (\ref{eq12}) we have a generally valid expression also for $\overRe(\gamma,\gamma_1,x_1)$.

In the kinetic equation (\ref{eq:kinel}) for the electron and positron densities $\npoel(p)$ the Compton terms can
be obtained by multiplying equation (\ref{eq18}) by $8\pi\lambdac^{-3}  p^3$. 

\subsection{Photon-photon pair production and pair annihilation}

The electron RKE accounting for 
pair-production and annihilation processes can be written as \citep{NL99}
\begin{equation} \label{eq:pp_el}
\fourpm \cdot \unb \noccm(\vecpm)= \frac{\re^2}{4} \frac{2}{\lambdac^3}
\int \frac{\rmd^3 \zpl}{\gpl} \frac{\rmd^3 x_1}{x_1} \frac{\rmd^3 x}{x}
\: \delta(\fourpm + \fourpp - \fourx_1 - \fourx)   \: \Fpp 
\left[ \noccx(\vecx_1) \noccx(\vecx) -  \noccm(\vecpm) \noccp(\vecpp) \right],
\end{equation}
where we used subscripts $\mp$ to explicitly show the momenta and the occupation number 
of electrons and positrons. 
Assuming homogeneity, we get 
\begin{equation} \label{pp1}
\left. \frac{D\noccm(\vecpm)}{Dt}  \right|_{\rm coll, pp}=
- c \: \sigmat \: \overline{s}_{\rm pa}(\vecpm) \: \Npos \: \noccm(\vecpm)
+ c \: \sigmat \: \Nphtwo \: \frac{\lambdac^3}{2} \: j_{\rm pp}(\vecpm),										
\end{equation}
where the pair annihilation cross-section (in units of $\sigmat$)  is given by
\begin{equation} \label{pp2}
\overline{s}_{\rm pa}(\vecpm) = \frac{3}{32\pi} \frac{2}{\lambdac^3\Npos} \frac{1}{\gmn}
\int \frac{\rmd^3 \zpl}{\gpl} \frac{\rmd^3 x_1}{x_1} \frac{\rmd^3 x}{x} \, \noccp(\vecpp) \: \Fpp
\: \delta(\fourpm + \fourpp - \fourx_1 - \fourx) 
\end{equation}
and the pair production rate by
\begin{equation}\label{pp3}
j_{\rm pp}(\vecpm) = \frac{3}{32\pi} \left(\frac{2}{\lambdac^3\Nph}\right)^2  \frac{1}{\gmn}
\int \frac{\rmd^3 \zpl}{\gpl} \frac{\rmd^3 x_1}{x_1} \frac{\rmd^3 x}{x} \, \noccx(\vecx) \: \noccpr(\vecx_1) \: \Fpp
\: \delta(\fourpm + \fourpp - \fourx_1 - \fourx).									
\end{equation}
The relativistically invariant  reaction rate $\Fpp$ is  \citep{LLVol4}
\begin{equation}\label{pp4}
\Fpp =  \frac{\xi}{\xi_1} + \frac{\xi_1}{\xi}  + 2 \; \left( \frac{1}{\xi} + \frac{1}{\xi_1}\right)
- \left( \frac{1}{\xi} + \frac{1}{\xi_1}\right)^2,									
\end{equation}
where $\xi = \fourpm\cdot\fourx= \fourpp\cdot\fourx_1$ and $\xi_1 =  \fourpm\cdot\fourx_1=  \fourpp\cdot\fourx$. 

Assuming again isotropic particle distributions  in frame $E$, we can write equation (\ref{pp3})  as
\begin{equation}\label{eq38_3}  
j_{\rm pp}(\zmn) = 3\pi  \left(\frac{2}{\lambdac^3\Nph}\right)^2 
\frac{1}{\gmn\zmn} \int_{x^{(L)}}^{\infty} \noccx(x) \: \rmd x  \int_{x_1^{(L)}}^{\infty}
\noccpr(x_1) \: \rmd x_1 \: \Rgg(\gmn,x,x_1) , 
\end{equation}
where we have defined
\begin{equation}\label{pp7}
\Rgg(\gmn,x,x_1) = 
\frac{1}{2} \frac{1}{(4\pi)^2}  x x_1 \zmn \int \frac{\rmd^3 \zpl}{\gpl} \: \rmd^2\omega \: \rmd^2\omega_1 \: \Fpp \:
\delta(\fourpm + \fourpp - \fourx_1 - \fourx).										
\end{equation}
The cross-section becomes
\begin{equation}\label{pp8}
\overline{s}_{\rm pa}(\zmn) = 4\pi \frac{2}{\lambdac^3\Npos} \int_0^{\infty} \zpl^2 \rmd\zpl  \: \sigmapa(\gpl,\gmn) \: \noccp(\zpl),	
\end{equation}
where
\begin{equation}\label{pp9}
\sigmapa(\gpl,\gmn) = \frac{3}{8} \frac{1}{(4\pi)^2}
\frac{1}{\gmn\gpl} \int \frac{\rmd^3 x_1}{x_1} \frac{\rmd^3 x}{x} \, \rmd^2\Opl \: \Fpp
\: \delta(\fourpm + \fourpp - \fourx_1 - \fourx).									
\end{equation}
The treatment of positrons is identical if we switch
the subscripts $-$ and $+$
in equations (\ref{eq:pp_el})--(\ref{pp9}).

The photon kinetic equation accounting for pair production/annihilation processes is
\begin{equation} \label{eq:pp_ph}
\fourx \cdot \unb \noccx(\vecx) = \frac{\re^2}{2} \frac{2}{\lambdac^3}
\int \frac{\rmd^3 x_1}{x_1} \frac{\rmd^3 \zmn}{\gmn} \frac{\rmd^3 \zpl}{\gpl} 
\: \delta(\fourpm + \fourpp - \fourx_1 - \fourx)   \: \Fpp 
\left[  \noccm(\vecpm) \noccp(\vecpp) -  \noccx(\vecx_1) \noccx(\vecx) \right] .
\end{equation}
Neglecting the spatial derivatives in the left hand side, this becomes
\begin{equation} \label{pp10}
\left.\frac{D \noccx(\vecx)}{Dt} \right|_{\rm coll, pp} =
- c \: \sigmat \: \overline{s}_{\rm pp}(\vecx) \: \Nph \: \noccx(\vecx)
+ c \: \sigmat \: \Nelmi \: \Npos \: \frac{\lambdac^3}{2}\ j_{\rm pa}(\vecx),
\end{equation}
where the pair-production cross-section is
\begin{equation}\label{pp11}
\overline{s}_{\rm pp}(\vecx) = \frac{3}{16\pi} \frac{2}{\lambdac^3\Nph} \frac{1}{x}
\int \frac{\rmd^3 x_1}{x_1} \frac{\rmd^3 \zmn}{\gmn} \frac{\rmd^3 \zpl}{\gpl} \, \noccpr(\vecx_1) \: \Fpp
\: \delta(\fourpm + \fourpp - \fourx_1 - \fourx)									
\end{equation}
and the emissivity due to pair annihilation
\begin{equation}\label{pp12}
j_{\rm pa}(\vecx) = \frac{3}{16\pi} \left(\frac{2}{\lambdac^3}\right)^2 \frac{1}{\Nelmi \Npos} \frac{1}{x}
\int \frac{\rmd^3 x_1}{x_1} \frac{\rmd^3 \zmn}{\gmn} \frac{\rmd^3 \zpl}{\gpl} \, \noccm(\vecpm) \: \noccp(\vecpp) \: \Fpp
\: \delta(\fourpm + \fourpp - \fourx_1 - \fourx).									
\end{equation}
Notice that unlike the electron equation, the photon equation is nonlinear owing to the fact that the cross-section (\ref{pp11}) depends explicitly on the photon distribution.

Under the isotropy assumption equations (\ref{pp11}) and (\ref{pp12})  in frame $E$ become
\begin{equation}\label{eq38_11}  
j_{\rm pa}(x) = 6\pi \,  \left(\frac{2}{\lambdac^3}\right)^2 \frac{1}{\Nelmi \Npos} \frac{1}{x^2} \int_{\gpl^{(L)}}^{\infty}
\noccp(\zpl) \: \rmd\gpl \: \int_{\gmn^{(L)}}^{\infty} \noccm(\zmn) \: \rmd\gmn  \Rgg(\gmn,x,x_1) , 
\end{equation}
where we have to substitute $x_1 = \gmn + \gpl - x$ from the energy conservation condition, and
\begin{equation}\label{pp14}
\overline{s}_{\rm pp}(x) = 4\pi \frac{2}{\lambdac^3\Nph} \int_{1/x}^{\infty} x_1^2 \rmd x_1 \: \sigmapp(x,x_1) \: \noccpr(x_1),
\end{equation}
where
\begin{equation}\label{pp15}
\sigmapp(x,x_1) = \frac{3}{4} \frac{1}{(4\pi)^2} 
\frac{1}{x x_1} \int \frac{\rmd^3 \zmn}{\gmn} \frac{\rmd^3 \zpl}{\gpl} \: \rmd^2 \omega_1
 \: \Fpp \: \delta(\fourpm + \fourpp - \fourx_1 - \fourx).							
\end{equation}
Explicit expressions for the rate $\Rgg(\gmn,x,x_1)$ (derived by \citealt{Sve82a}, see also \citealt{BS97} and \citealt{NL99}) 
and the cross-sections $\sigmapa(\gpl,\gmn)$, $\sigmapp(x,x_1)$ as well as the lower integration limits in equations (\ref{eq38_3}) and (\ref{eq38_11}) are given in Appendix \ref{app:pairs}.

The pair-production terms in equations (\ref{eq:kinph}) and (\ref{eq:kinel}) take the form
\beq \label{eq38_1}
\dotn_{\rm ph,pp} (x)  & = & -c \, \alpha_{\rm pp}(x) \, \nph(x) + \epsilon_{\rm pa}(x)	, \\	
\label{eq38_2}
\dotn_{\rm \pm,pp} (\zplmn)  & = & -c \, \alpha_{\rm pa}(\zplmn) \, \npoel(\zplmn) + \epsilon_{\rm pp}(\zplmn).
\eeq
By comparing with equations (\ref{pp1}) and (\ref{pp10}) we find the absorption coefficients and emissivities to be
\beq
\alpha_{\rm pp}(x) & =&  \sigmat \: \overline{s}_{\rm pp}(x) \: \Nph,  \qquad 
\epsilon_{\rm pa}(x) = 4\pi \: c \: \sigmat \: \Nelmi \: \Npos \: x^3 \: j_{\rm pa}(x) , \\
\alpha_{\rm pa}(\zplmn) & =&  \sigmat \: \overline{s}_{\rm pa}(\zplmn) \: \Nmnpl, \qquad
\epsilon_{\rm pp}(\zplmn) = 4\pi \: c \: \sigmat \: \Nphtwo \: \zplmn^3 \: j_{\rm pp}(\zplmn).
\eeq

\subsection{Synchrotron radiation}

The kinetic equations describing synchrotron radiation need to be written in frame $E$, where we assume there is only tangled magnetic field (and no electric field). Using the Einstein coefficients and the cross-sections describing synchrotron emission and absorption \citep{GS91}, we get the collision terms for these processes in the
electron/positron and photon equations \citep[see also][]{ochel79}: 
\beq 	\label{eq:syn:el_kin}
\left. \frac{D}{Dt}  \left[\gamma p \, \noccpm(p)\right] \right|_{\rm coll, syn} &=& 
\int_0^{\infty} \rmd x \int_{\gamma}^{\infty} \rmd\gamma_1 \: \gamma_1 p_1 \: \frac{P(x,\gamma_1)}{x} \: \delta(\gamma_1 - \gamma - x) \:
\left\{  \noccpm(p_1) \: [1 + \noccx(x)] - \noccpm(p) \: \noccx(x)
\right\}		\nonumber \\
&-& \int_0^{\infty} \rmd x \int_{1}^{\gamma} \rmd\gamma_1 \: \gamma p \: \frac{P(x,\gamma)}{x} \: \delta(\gamma - \gamma_1 - x) \: \left\{ \noccpm(p) \: [1 + \noccx(x)] - \noccpm(p_1) \: \noccx(x) \right\} ,
\eeq
\begin{equation}\label{eq:syn:ph_kin}
\left.\frac{D}{Dt} \left[ x^2 \noccx(x) \right]\right|_{\rm coll, syn} =
\int_{1}^{\infty} \rmd\gamma \int_{1}^{\gamma} \rmd\gamma_1 \: \gamma p \: \frac{P(x,\gamma) }{x} \:
\delta(\gamma - \gamma_1 - x) \: \left\{ \nocce(p)[1+\noccx(x)] - \nocce(p_1) \noccx(x) \right\} .
\end{equation}
Here $P(x,p)$ is the angle-integrated cyclo-synchrotron spectrum of a single electron, normalized  
to the electron cooling rate: 
\begin{equation} \label{eq:gammas}
\int_{0}^{\infty} P(x, \gamma) \, \rmd x = - \dotgammas = \frac{4 }{3} \frac{\sigmat \UB}{\me c} p^2  ,
\end{equation}
where $\UB=B^2/(8\pi)$ is the magnetic energy density. One can readily verify that equations (\ref{eq:syn:el_kin})
conserve the total number of electrons and positrons, and that the total energy is conserved by equations (\ref{eq:syn:el_kin}) and (\ref{eq:syn:ph_kin}).

Under the physical conditions that we are interested in, the average energy (or momentum) of
an emitted or absorbed photon is much lower than the energy (momentum) of the electron taking part in the process.
The standard way is therefore to treat synchrotron processes as continuous cooling or heating for electrons
and as an emission or absorption process for photons.

We write the photon terms in the form
\begin{equation} \label{eq23}
\left.\frac{D \noccx(x)}{Dt} \right| _{\rm coll, syn} = 
-c \, \alphas(x) \, \noccx(x) + \frac{\lambdac^3}{8\pi} \frac{\epssyn(x)}{x^3} , 
\end{equation}
where $\alphas$ and $\epssyn$ 
are cyclo-synchrotron absorption and emission coefficients, respectively. 
In the kinetic equation (\ref{eq:kinph}) for the photon density $\nph(x)$ the corresponding term can
be obtained by multiplying equation (\ref{eq23}) by $8\pi\lambdac^{-3} x^3$: 
\begin{equation} \label{eq901}
\dotn_{\rm ph,syn} (x) =  -c \, \alphas(x) \, \nph(x) + \epssyn(x) . 
\end{equation}
The emissivity $\epssyn$ 
gives the number of photons emitted per logarithmic dimensionless energy interval $\rmd\ln{x}$,
per unit volume and time 
and can be identified by comparing the corresponding terms in 
equations (\ref{eq:syn:ph_kin}) and (\ref{eq23}): 
\begin{equation} \label{eq33}
\epssyn(x) =  \frac{8\pi}{\lambdac^3}
\int P(x,\gamma) \; p^2 \nocce(p) \; \rmd p = 
\int P(x,\gamma) \; \nel(p) \; \rmd \ln{p} . 
\end{equation}
Similarly, by comparing the terms proportional to $\noccx$ we identify the absorption coefficient 
\citep[e.g.][]{RL79}: 
\begin{equation} \label{eq35}
\alphas(x) = \frac{1}{4\pi c x^3}  \int \rmd ^3p \left[ \nocce(p_1) - \nocce(p) \right] P(x, \gamma) = 
- \frac{1}{c \ x^2}  \int    \frac{\rmd \nocce(p)}{\rmd p} P(x,\gamma) \: \gamma  p \ \rmd p\ ,
\end{equation}
where  $p_1=\sqrt{(\gamma - x)^2-1}$ is the electron momentum corresponding to energy $\gamma_1 = \gamma - x$ and 
the second expression is obtained by expansion  to the first order in $x \ll \gamma$. 

In terms of the electron number density $\nel(p)$ the absorption coefficient takes the form: 
\begin{equation} \label{eq35a}
\alphas(x)  = \frac{\lambdac^3}{8\pi c}  
\frac{1}{x^2} \int \frac{\gamma P(x,\gamma)}{p^2} \left[ 3 \nel(p)
- \frac{\rmd \nel(p)}{\rmd\ln{p}} \right] \: \rmd\ln{p}. 				
\end{equation}

The synchrotron processes for electrons can be treated as continuous using
the Fokker-Planck equation. It can be obtained from equation (\ref{eq:syn:el_kin})
employing the delta-function to take the
integral over $\gamma_1$ and expanding $\gamma_1 p_1 \ P(x,\gamma_1)$ and $\npoel(p_1)$ near $p$
to the second order in the small 'parameter' $x$. Collecting the terms and finally integrating over the photon energy $x$
we get
\begin{equation} \label{eq36}
\frac{\partial}{\partial t} \left[\gamma p \, \noccpm(p)\right] = -\frac{\partial}{\partial \gamma}
\left[ \dotgammas \ \gamma  p\ \noccpm(p) - H(p)\ \gamma p\ \frac{\partial \noccpm(p)}{\partial \gamma}  \right]
+\frac{1}{2} \, \frac{\partial^2}{\partial\gamma^2} \left[  H_0(p) \,\gamma p \, \noccpm(p)
\right],
\end{equation}
where
\begin{equation} \label{eq37}
H(p) =  \int  P(x,\gamma) \ \noccx(x) \; x \ \rmd x  = \frac{\lambdac^3}{8\pi} \int  \frac{P(x,\gamma) }{x} \ \nph(x) \ \rmd \ln{x}, 
 \qquad H_0(p) =  \int  P(x,\gamma)  \; x \ \rmd x.	
\end{equation}
To get the total electron energy gain/loss rate, one has to multiply equation~(\ref{eq36}) by $8\pi \lambdac^{-3}\gamma \, \rmd \gamma$ and integrate.
Multiplying equation~(\ref{eq23}) by
$ 8\pi \lambdac^{-3}x^3 \rmd x$ and integrating gives the corresponding rate for photons. Using expressions (\ref{eq:gammas}), (\ref{eq33}), (\ref{eq35}) and (\ref{eq37}), we can verify that energy conservation is maintained when switching from equation (\ref{eq:syn:el_kin}) to the continuous approximation (\ref{eq36}).

Note that the last term on the rhs of equation (\ref{eq36})  is missing in similar equations derived previously \citep{McCray69,GGS88}. It corresponds to the diffusion due to spontaneous emission, but does not contribute to the electron cooling/heating. However, in most cases we expect its contribution to be negligible compared to the other terms. It is of the order $x/\gamma$ smaller than the cooling term with $|\dotgammas|$ and, when electrons are mildly-relativistic, self-absorption becomes important, $\noccx\gg 1$ and the term containing $H$ dominates. Therefore, we neglect the term with $H_0$ in our simulations. Thus, for the distributions $\npoel(p)$, equation~(\ref{eq36}) takes the form
\begin{equation} \label{eq:syndiff}
\dotn_{\rm \pm,syn} (p)  = -\frac{\partial}{\partial \ln{p}} 
\left[ A_{\rm e,syn}(p) \npoel(p) 
	- B_{\rm e,syn}(p) \frac{\partial \npoel (p)}{\partial\ln{p}} \right] ,
\end{equation}
where 
\beq \label{eq:AB_elsyn}
A_{\rm e,syn}(p)  =  \left( \dotgammas + 3 \frac{\gamma}{p^2}H(p) \right) \;\frac{\gamma}{p^2} , \qquad 
B_{\rm e,syn}(p)  =  H(p)\frac{\gamma^2}{p^4}  .
\eeq

It is worth mentioning here that other emission/absorption processes, e.g. bremsstrahlung, can be implemented 
analogously to the synchrotron radiation, once the emissivity function of a single electron $P(x,\gamma)$ (which now 
may depend on the particle distribution) is specified. 

\subsection{Coulomb collisions}

The RKE accounting for electron (positron) evolution due to Coulomb 
scatterings is
\begin{equation} \label{eq:cc_el}
\fourp \cdot \unb \noccpm(\vecp)= \re^2 \frac{2}{\lambdac^3}
\int \frac{\rmd^3 p_1}{\gamma_1} \frac{\rmd^3 p'_1}{\gamma'_1} \frac{\rmd^3 p'}{\gamma'}
\: \delta(\fourp_1 + \fourp - \fourponeprime - \fourpprime )   \: \Fcc 
\left[ \nocce(\vecponeprime) \noccpm(\vecpprime) - \nocce(\vecp_1) \noccpm(\vecp)  \right] .
\end{equation}
The invariant reaction rate for M{\o}ller scattering (i.e. $e^-e^-$ and $e^+e^+$) is given by \citep{LLVol4} 
\begin{equation}
\Fcc = \left( \frac{\xi_1}{\xi - 1} + \frac{\xi}{\xi_1 - 1} + 1 \right)^2
+\frac{1 - 4\xi\xi_1}{(\xi - 1)(\xi_1 - 1)} + 4
\end{equation}
and the scalar products of particles' four-momenta are defined as $\xi = \fourp\cdot\fourp'$ and $\xi_1 =  \fourp_1\cdot\fourp'$.
As discussed by \citet{Bar87} and \citet{CB90}, the corresponding rates  for  Bhabha $e^\pm e^\mp$ scattering 
 are nearly the same in the small-angle scattering approximation, we therefore do not distinguish between electrons and positrons 
 in these equations.  

Although the Coulomb process is collisional in nature, it is customary to treat it in the Fokker-Planck framework, i.e. as a continuous diffusive energy exchange mechanism. This is due to the well-known divergent nature of the Coulomb cross-section for small-angle scatterings with negligible energy exchange per event, while in the parameter regimes we are interested in, a large number of such scatterings dominates the energy gain or loss rate of a particle over a much smaller number of large-angle scatterings. 
In  frame $E$, where the particle distributions are approximately homogeneous and isotropic, we can
therefore write the Coulomb terms in the form of the 
Fokker-Planck equation in (scalar) momentum space
\begin{equation} \label{Coulnum:eq1}
\dotn_{\rm \pm,Coul} (p)  = -\frac{\partial}{\partial \ln{p}} 
\left[ A_{\rm e,Coul}(p) \npoel(p) 
	- B_{\rm e,Coul}(p) \frac{\partial \npoel (p)}{\partial\ln{p}} \right] 
\end{equation} 
with coefficients given by 
\begin{equation} \label{Coul:AB}
A_{\rm e,Coul}(p) =   \frac{\dotgammaCoul \gamma}{p^2} 
- \frac{\partial}{\partial\gamma} \left( \frac{1}{2} \frac{\gamma \, \DCoul}{p^2} \right)  , 
\qquad B_{\rm e,Coul}(p)  =  \frac{1}{2} \frac{\gamma^2 \DCoul}{p^4}.
\end{equation}
The energy exchange rate and the diffusion coefficient can be obtained by calculating the first and second moments of equation~(\ref{eq:cc_el}) keeping only small-angle scatterings and are expressed as integrals over the particle distributions:
\begin{equation} 
\dotgammaCoul = \int a(\gamma,\gamma_1) \, \nel(p_1) \, \rmd\ln{p_1}, \qquad \DCoul(p) = \int d(\gamma,\gamma_1) \, \nel(p_1) \, \rmd\ln{p_1}.
\end{equation} 
The rates $a(\gamma,\gamma_1)$ and $d(\gamma,\gamma_1)$ have been calculated by \citet{NM98} and
are given in Appendix \ref{app:Coul}.

\section{Numerical treatment}

We numerically solve the set of coupled integro-differential equations of the general form 
(\ref{eq:rke_gen_ph})--(\ref{eq:rke_gen_el}).
We first define an equally spaced grid in the logarithms of particles' momenta:
\beq
\ln{p}_i & =&  \ln{p}_{\min} + (i-1) \cdot \gridp, \quad i \in [1,i_m], \\
\ln{x}_l & = & \ln{x}_{\min} + (l-1) \cdot \gridx, \quad l \in [1,l_m].
\eeq
Writing all differentials and integrals on the finite grids, we get three systems (for photons, electrons and positrons)
 of linear algebraic equations of the general form
\begin{equation}
\frac{\distr_{i}^{k+1} - \distr_{i}^{k}}{\Delta t_k} = \sum_{i^{\prime} = 1}^{i_m} M^{k+1/2}_{i,i^{\prime}} \cdot\frac{1}{2}
\left( \distr ^{k+1}_{i^{\prime}} + \distr ^{k}_{i^{\prime}}     \right),
\end{equation}
where $\Delta t_k$ is the size of the $k$-th (variable) timestep. Such semi-implicit differencing scheme, where both sides
of the equation are centered at timestep $k+1/2$, is known as the Crank-Nicolson scheme \citep[see e.g.][]{NumRec92}.
All physics is contained within the matrix $M_{i,i^{\prime}}$, which can be explicitly calculated at each step.
The systems of equations for all types of particles are solved stepwise, alternating between equations and
requiring a matrix inversion at every step.
After solving a set of equations for photons, 
the updated photon distribution is used to calculate matrix $M$ for electron and positron equations. 
Then we solve for distributions of electrons/positrons and substitute it to the photon equation and so on.

\subsection{The Chang and Cooper scheme}

The matrix $M_{i,i^{\prime}}$ of the linear system can be decomposed into two parts arising from the differential
and integral terms in equations (\ref{eq:rke_gen_ph})--(\ref{eq:rke_gen_el}). 
The differential part contributes a tridiagonal matrix, the form of the equation (e.g. for electrons), 
giving rise to it, is
\begin{equation}\label{eq42}
\frac{\distr_{i}^{k+1} - \distr_{i}^{k}}{\Delta t_k} =
 -\frac{1}{\gridp} \left[ F^{k+1/2}_{i+1/2} - F^{k+1/2}_{i-1/2} \right],									
\end{equation}
where the momentum space flux is given by
\begin{equation}\label{eq43}
F^{k+1/2}_{i+1/2} = A^{k+1/2}_{i+1/2} \; \distr_{i+1/2}^{k+1/2} - B^{k+1/2}_{i+1/2} \;
\frac{\distr_{i+1}^{k+1/2} - \distr_{i}^{k+1/2}}{\gridp}.				
\end{equation}
The distribution function between time gridpoints is defined according to the Crank-Nicolson scheme as (omitting the momentum index)
\begin{equation}\label{eq44}
\distr^{k+1/2} = \frac{1}{2} \left( \distr^{k+1} + \distr^{k}   \right).				
\end{equation}
We also have to somehow define the distribution function between momentum gridpoints. Following \citet{CC70} we introduce a parameter $\delta_i$ so that (now omitting the time index)
\begin{equation}\label{eq45}
\distr_{i+1/2} = (1-\delta_i) \distr_{i+1} + \delta_i \distr_{i}, \quad \delta_i \in [0,1].	
\end{equation}
The basic idea of the Chang and Cooper scheme is to employ this parameter to ensure that the differencing scheme converges
to the correct equilibrium solution independently of the size of the gridstep $\gridp$.
Assuming that the momentum space flux through the boundaries vanishes, the equilibrium solution tells us that
it must vanish everywhere, i.e. $F = 0$. From equations (\ref{eq43}) and (\ref{eq45}) we then have
\begin{equation}\label{eq46}
\frac{\distr_{i+1}}{\distr_{i}} = \frac{\delta_i \, A_{i+1/2} \,
\gridp + B_{i+1/2}}{B_{i+1/2} - (1-\delta_i) \, A_{i+1/2} \, \gridp},	
\end{equation}
while the exact solution gives \citep{CC70}
\begin{equation}\label{eq47}
\frac{\distr_{i+1}}{\distr_{i}} = \exp \left[\frac{A_{i+1/2}}{B_{i+1/2}} \gridp \right].	
\end{equation}
We can see that using either centered-differencing $(\delta = 1/2)$ or forward differencing $\delta = 0$,
equations (\ref{eq46}) and (\ref{eq47}) agree only to the first order in $A\,\gridp/B$.
To make the correspondence exact, one has to equate the two equations and solve for $\delta_i$, to get
\begin{equation}
\delta_i = \frac{1}{w_i} - \frac{1}{\exp(w_i) - 1}, \qquad w_i = -\frac{A_{i+1/2}}{B_{i+1/2}} \, \gridp. 
\end{equation}
Aside from converging to the correct equilibrium solution, such choice of $\delta_i$ also
guarantees positive spectra, as shown by \citet{CC70}. 
Although this method applies to purely differential equations, we can still use it in our
integro-differential equations to ensure that the differential part {\it tends} toward its own
correct equilibrium solution, which would also be the correct solution for the full equation in the region
where the differential terms happen to dominate.

\subsection{Treatment of Compton scattering}

Accurate numerical treatment of Compton scattering over a wide range of energies is not straightforward. This is caused by the well-known fact that at different energies the process takes place in different regimes. If the energy of a photon in electron rest frame  is much smaller than the electron rest energy, the process takes place in the Thomson regime and the electron loses a very small amount of its energy in one scattering. Correspondingly, there is a sharp peak in the electron redistribution function $\Rel$ near $p = p_1$. We cannot therefore numerically resolve $\Rel$ on our finite grid and have to treat the energy loss process as continuous. On the other hand, for scattering in the  Klein-Nishina regime the electron can lose a significant amount of its energy in one scattering. Wishing to include both regimes, we need a way to switch from the continuous approximation (implying a differential equation) to direct calculation of scattering through the integral terms.   Similar treatment is required for photons, although the continuous approximation is only needed in the regime where the photon energy is much lower than the electron rest energy and the electron is non-relativistic.

\subsubsection{Scattering of electrons: separation of regimes}

Let us first look at the electron redistribution function (\ref{eq19}). We wish to know what is the lowest incoming photon energy $ x_{\star}^{\pm}(p_1)$ that can cause a shift in electron momentum $p_1$ by $|\Delta\ln{p}|$.  This energy is related to the lower limit (\ref{eq21}) of the integral in equation~(\ref{eq19}). If the shift is small enough, we can write
\begin{equation}\label{eq60}
 x_{\star}^{\pm}(p_1)  \approx  x_{\star}(\gamma,\gamma_1)= \frac{1}{2}  \left( \pm |\Delta\gamma| + |\Delta p|
\right) \approx \frac{1}{2} p_1 \, |\Delta\ln{p_1}| \left(	1 \pm \frac{p_1}{\gamma_1}	\right) ,	
\end{equation}
where we have used $p\,\rmd p = \gamma \, \rmd\gamma$. The plus sign applies when the electron gains energy and  the minus when it loses it. We see that for high energy electrons, the minimum energy of photons for which we can resolve up- or downscattering
is vastly different. However, since the upscattering (energy increase) of relativistic electrons is extremely inefficient, we concern ourselves only with being able to resolve their downscattering (i.e. cooling) and so use the minus sign in equation~(\ref{eq60}). Choosing $|\Delta\ln{p_1}|$ comparable to our grid step (we use somewhat arbitrarily $3\gridp$) in the electron equation, we then state that scattering of electrons on photons with $x_1 < x_{\star}^{-}(p_1)$ cannot be resolved.

We now split the redistribution function into two parts according to whether we can or cannot resolve it on our grid
\begin{equation}\label{eq63}
\Se(p,p_1) = \Se^{<}(p,p_1) + \Se^{>}(p,p_1),
\end{equation}
where for the first term the integral in equation~(\ref{eq19}) is taken over  $x_1 < x_{\star}^{-}(p_1)$, 
and the second term is defined by integrating over the remaining $x_1$. 
To totally isolate scatterings that undergo on photons with energies below and above $x_{\star}^{-}$,
we have to write the extinction coefficient as an analogous sum,
$\overline{s}_0(p) = \overline{s}_0^{<}(p) + \overline{s}_0^{>}(p)$, where
\begin{equation}\label{eq66}
\overline{s}^{\lessgtr}_0(p) = \frac{4\pi}{\gamma} \int \Se^{\lessgtr}(p_1,p)\ p_1 \, \rmd\gamma_1 \; , 
\end{equation}
in accordance with equation~(\ref{eq22a}). For the terms containing $\Se^{>}$ and $\overline{s}_0^{>}$ in the electron equation, we compute the integrals through the discrete sums, but the terms containing $\Se^{<}$ and $\overline{s}_0^{<}$
have to be accounted for by continuous energy exchange terms in the equation. Since we also want to treat thermalization  by Compton scattering, these terms have to contain a second order derivative of the electron distribution (a diffusive term). 
Therefore, we take the  standard form of the Fokker-Planck equation 
\begin{equation}\label{eq67}
\dotN_{\rm \pm,diff,cs}(\gamma) = 
-\frac{\partial}{\partial\gamma} \left\{ \dotgammac \, \Nplmn(\gamma) - 
\frac{1}{2} \frac{\partial}{\partial\gamma} \left[ \De(\gamma) \, \Nplmn(\gamma) \right] \right\},		
\end{equation}
while the exact equation for the ($<$) terms comes from equation~(\ref{eq18}), written here for $\Nplmn(\gamma)$
\begin{equation}\label{eq68}
\dotN_{\rm \pm,coll,cs}^{<} (\gamma)
= -c \, \sigmat  \, \overline{s}_0^{<}(p)\; \Nph \,  \Nplmn(\gamma)
+ 4\pi \, c \, \sigmat \Nph \, p \int \frac{\rmd\gamma_1}{\gamma_1} \: \Se^{<}(p,p_1) \: \Nplmn(\gamma_1). 			
\end{equation}
In order to make a physically sensible correspondence between these two representations, we
demand that the first three moments of equations (\ref{eq67}) and (\ref{eq68}) were identical.
Substituting equation~(\ref{eq66}) to (\ref{eq68}) we find
\beq \label{eq69}
\int \dotN_{\rm \pm,coll,cs}^{<}(\gamma) \gamma^i \rmd\gamma =
4\pi \, c \, \sigmat \Nph \, \int \rmd\gamma  \int \rmd\gamma_1 \, \gamma^i
\left\{
-\frac{p_1}{\gamma} \, \Se^{<}(p_1,p) \, \Nplmn(\gamma) + \frac{p}{\gamma_1} \, \Se^{<}(p,p_1) \, \Nplmn(\gamma_1)
\right\}									\nonumber \\ 
= 4\pi \, c \, \sigmat \Nph \, \int \rmd\gamma \int \rmd\gamma_1 \frac{p_1}{\gamma} \left( \gamma_1^i - \gamma^i \right) \,
\Se^{<}(p_1,p) \, \Nplmn(\gamma) = 
c \, \sigmat \Nph \, \int \rmd\gamma \; \overline{(\gamma_1^i - \gamma^i)} \, \overline{s}^{<}_0(p) \,
\Nplmn(\gamma) \, , 
\eeq
where similarly to the moments of the photon redistribution function (NP94), 
we defined the moments of the electron redistribution function 
\begin{equation}\label{eq70}
\overline{\gamma_1^i} \, \overline{s}^{<}_0(p) 
\equiv \frac{4\pi}{\gamma} \int p_1  \gamma_1^i \, \rmd\gamma_1 \; \Se^{<}(p_1,p).	
\end{equation}
The zeroth moment (giving zero in the rhs of eq. [\ref{eq69}]) is just a statement of particle number conservation, while the first moment gives the total rate at which the electrons gain (or lose) energy. The moments defined by equation~(\ref{eq70}) can be calculated analytically using the exact Klein-Nishina scattering cross-section. For photons this was shown by NP94, while the extension of these calculations to the electrons is given in Appendix \ref{app:moments}.

The moments of the continuous approximation (\ref{eq67}) are
\beq
\int \dotN_{\rm \pm,diff,cs}  (\gamma) \,
\rmd\gamma & =&  0, \\
\int \dotN_{\rm \pm,diff,cs} (\gamma) \,
\gamma \, \rmd\gamma & =& 
\int \dotgammac \, \Nplmn(\gamma) \, \rmd\gamma , \\
\int \dotN_{\rm \pm,diff,cs} (\gamma) \,
\gamma^2 \, \rmd\gamma & =& 
\int \left[ 2\gamma \dotgammac + \De(\gamma) \right] \, \Nplmn(\gamma) \, \rmd\gamma.
\eeq
Here we have assumed that the distribution function $\Nplmn(\gamma)$ vanishes at the boundaries of integration.
Exact correspondence with equation~(\ref{eq69}) can be made if we identify
\begin{equation} \label{Cnum:gd}
\dotgammac  =  c \, \sigmat \Nph \, \overline{(\gamma_1 - \gamma)} \, \overline{s}^{<}_0(p) , \qquad 
\De(\gamma)  = c \, \sigmat \Nph \, \overline{(\gamma_1 - \gamma)^2} \, \overline{s}^{<}_0(p) ,
\end{equation}
while for the zeroth moment the correspondence is automatic.
These moments can be computed using equations (\ref{eq139}) and (\ref{eq140}).
Finally, we write equation~(\ref{eq67}) through $\npoel(p)$ and in the form that can be included in the Chang \& Cooper
differencing scheme together with other terms
\begin{equation}\label{eq77}
\dotn_{\rm \pm,diff,cs} (p)  = 
-\frac{\partial}{\partial \ln{p}} \left[ A_{\rm e,cs}(p) \npoel(p) 
	- B_{\rm e,cs}(p) \frac{\partial \npoel (p)}{\partial\ln{p}} \right] , 
\end{equation}
where 
\beq \label{eq:AB_el_cs}
A_{\rm e,cs}(p) =   \frac{\dotgammac \gamma}{p^2} 
- \frac{\partial}{\partial\gamma} \left( \frac{1}{2} \frac{\gamma \, \De(\gamma)}{p^2} \right)  , 
\qquad B_{\rm e,cs}(p)  =  \frac{1}{2} \frac{\gamma^2 \De(\gamma)}{p^4} . 
\eeq

\subsubsection{Scattering of photons and three-bin approximation}

Insufficient resolution of numerical calculations can become an issue also for the scattering of photons
if the electron energies are low enough.
A photon will then exchange very little energy with an electron upon scattering and the redistribution function
is strongly peaked near $x=x_1$.
To overcome this we propose the following approach. We separate scatterings that take place within
some narrow interval around the energy of the incoming photon from those invoking photon energy outside this interval.
We then approximate the scatterings taking place within the central interval by a continuous process
and account for this by differential terms calculated through the exact moments of the redistribution function.

To keep the correspondence to the electron equation, we rewrite the photon evolution equation (\ref{eq8})
in terms of $\Nph(x)$:  
\beq\label{eq101}
\dotN_{\rm ph,coll,cs} (x)  & = & 
4\pi \, c \, \sigmat \Nel 
\left \{
\int_{\notin}  \: \rmd {x_1}
\left[ \Nph(x_1) \frac{x}{x_1} \Sph(x,x_1) - \Nph(x) \frac{x_1}{x} \Sph(x_1,x)
\right] \right. 		\nonumber \\
&+& \left.
\int_\in \: \rmd  {x_1}
\left[ \Nph(x_1) \frac{x}{x_1} \Sph(x,x_1) - \Nph(x) \frac{x_1}{x} \Sph(x_1,x)
\right] \right\}, 							
\eeq
where  the extinction coefficient  is expressed explicitly through $\Sph$ using equation~(\ref{eq13a}). 
Here $\in$ stands for the interval $[x {\rm e}^{- \deltax},x {\rm e}^{\deltax}]$ and 
$\notin$ means integration  from $0$ to $\infty$ excluding that interval. 
The width of the central region ($2 \deltax$ in log units) is somewhat arbitrary, but should include at least a couple of bins,
with our choice being three, i.e. $\deltax=\frac{3}{2} \gridx$.

For the second integral in equation~(\ref{eq101}) we wish to write a continuous approximation similar to equation~(\ref{eq67})
\begin{equation}\label{eq101a}
\dotN_{\rm ph,diff,cs} (x)  = 
-\frac{\partial}{\partial x} \left\{ \dotxc \, \Nph(x) - \frac{1}{2} \frac{\partial}{\partial x} \left[ \Dph(x) \, \Nph(x)
\right] \right\}.					
\end{equation}
Similarly to what was done for electrons, the coefficients in equation~(\ref{eq101a}) are determined from the requirement that the first three moments of the differential and integral equations coincide. The moments of the 'central' part of equation~(\ref{eq101}) (denoted by $\in$) are
\begin{equation}\label{eq102}
\int_0^{\infty} \dotN_{\rm ph,coll,cs}^{\in} (x) \, x^i\, \rmd x 
= 4\pi \, c \, \sigmat \Nel \int_{0}^{\infty} \rmd {x} \int_{\in}
\rmd {x_1} \:  \left( x_1^i - x^i \right) \, \frac{x_1}{x}\, \Sph(x_1,x) \, \Nph(x), 					
\end{equation}
where  the integration limits for $x$ and $x_1$ in the first term were switched, 
because for constant $\deltax$ the area on the $({x},{x_1})$ plane is the same.
The moments of the differential equation are similar to what were obtained for electrons
\beq\label{eq102a}
\int_0^{\infty} \dotN_{\rm ph,diff,cs}^{\in} (x) \,
\rmd x & =&  0,	\\	
\label{eq102b} 
\int_0^{\infty} \dotN_{\rm ph,diff,cs}^{\in} (x) \,
x \, \rmd x & =& 
\int_0^{\infty} \dotxc \, \Nph(x) \, \rmd x	\, , 			\\	
\label{eq102c}
\int_0^{\infty} \dotN_{\rm ph,diff,cs}^{\in} (x) \,
x^2 \, \rmd x & = & 
\int_0^{\infty} \left[ 2x \dotxc + \Dph(x) \right] \, \Nph(x) \, \rmd x.		
\eeq
Equations (\ref{eq102}) and (\ref{eq102a})--(\ref{eq102c}) give identical expressions for the first three moments
of the 'central' part of the equation if we identify
\begin{equation}\label{eq103}
\dotxc   =  4\pi \, c \, \sigmat \Nel \int_{\in} \rmd{x_1} \:
\left( x_1 - x \right) \, \frac{x_1}{x} \, \Sph(x_1,x)	, \qquad
\Dph(x)  =  4\pi \, c \, \sigmat \Nel \int_{\in} \rmd {x_1} \:
\left( x_1 - x \right)^2 \, \frac{x_1}{x} \, \Sph(x_1,x).			
\end{equation}
The 0-th moment is identically zero for both equations~(\ref{eq102}) and (\ref{eq102a}), implying particle conservation.

The integrals in equations~(\ref{eq103}) are computed numerically at a finer grid. At low photon energies, the redistribution function can be narrower than the whole integration interval, and integration can present a problem. In this case, however,  
we can extend the integration limits in equations~(\ref{eq103}) from $0$ to $\infty$ and to calculate the moments of the redistribution function analytically (NP94). Using the limits on $\gamma_{\star}$, given by equation~(\ref{eq13}), one can show that scattering takes place entirely within the central interval $\in$ for incident photons and electrons satisfying the following relations:
\begin{equation}\label{eq111}
x < \frac{\deltax}{2}, \qquad 
p < p_{\star}^{-}(x) = \frac{\deltax}{2} - x.	
\end{equation}
We can write the moments of the redistribution function in a way similar to equation~(\ref{eq70}):
\begin{equation} \label{eq113a}
\overline{x_1^i} \, \overline{s}_0^{<}(x)  \equiv
\frac{4\pi}{x} \int x_1^{i+1} \, \rmd x_1 \; \Sph^{<}(x_1,x),										
\end{equation}
where the $<$ superscript signifies that only  electrons with $p < p_{\star}^{-}(x)$ are taken into account. 
Equations (\ref{eq103}) then (for  $x< \deltax/2$) become
\begin{equation}\label{eq113b}
\dotxc  =  c \, \sigmat \Nel\ \overline{(x_1-x)} \, \overline{s}_0^{<}(x) , \qquad
\Dph(x) =  c \, \sigmat \Nel\ \overline{(x_1-x)^2} \, \overline{s}_0^{<}(x) ,
\end{equation}
and can be
computed using equations (\ref{eq133})--(\ref{eq134}).

For numerical differencing equation~(\ref{eq101a}) has to be written in the form
\begin{equation}\label{eq:diff_ph_cs}
\dotn_{\rm ph,diff,cs}(x) = 
-\frac{\partial}{\partial \ln{x}} \left[ A_{\rm ph,cs}(x) \nph(x) 
	- B_{\rm ph,cs}(x) \frac{\partial \nph (x)}{\partial\ln{x}} \right] , 
\end{equation}
where 
\beq \label{eq:AB_ph_cs}
A_{\rm ph,cs}(x)  =   \frac{\dotxc}{x} - \frac{\partial}{\partial x} \left( \frac{1}{2} \frac{\Dph(x)}{x} \right) ,  
\qquad B_{\rm ph,cs}(x) =  \frac{1}{2} \frac{\Dph(x)}{x^2} . 
\eeq

\subsection{Pair production and annihilation}

The numerical treatment of pair-production and annihilation processes in our code is fairly straightforward. The only potential difficulty can arise from the non-linearity of the absorption term in the photon equation. To deal with this we have chosen the simplest possible approach: for calculating the pair-production opacity at each step we simply use the photon distribution from the previous step. The error caused by doing so is not expected to be significant in most cases. It is well-known that a photon of energy $x$ will most efficiently interact with photons of energy $x_1 \approx 3/x$, thus if its energy is not very close to the electron/positron rest energy, the photon will most likely annihilate on another photon of a vastly different energy than its own. Therefore, we can visualize two separate populations of photons that pair-produce on each other, with the dividing energy at $\me c^2$. The photon distribution from the previous step is then taken to be the 'target'  population on which the photons that are being evolved pair produce.

Since we wish the numerical scheme to treat electrons and positrons identically (particularly when we are dealing with pure pair plasma), while at each step one of them has to be evolved first when the outcome of the other is yet unknown, we use a fully implicit scheme for the pair annihilation terms.

The quantities $\Rgg(\gmn,x,x_1)$ 
$\sigmapa(\gpl,\gmn)$ and $\sigmapp(x,x_1)$ defined by equations (\ref{pp7}), (\ref{pp9}) and (\ref{pp15}) are precalculated on a fine grid and thereafter averaged within the electron/positron and photon bins used by the code.
The integrals in the expressions (\ref{eq38_3}), (\ref{pp8}), (\ref{eq38_11}) and (\ref{pp14}) for emissivities and absorption coefficients are calculated through discrete sums.

\subsection{Treatment of synchrotron processes}

One of the main difficulties in numerically treating synchrotron processes in compact sources is that
the optical thickness of the medium due to self-absorption might become extremely large
at low energies compared to, say, Thomson optical thickness.
Almost all photons that are produced are immediately absorbed, so very few escape.
But the energy which those few carry away comes from the small net energy exchange rate between electrons and photons,
which we need to keep track of to maintain the energy balance. Near the equilibrium,  in the photon equation we have
two large terms describing emission and absorption, which nearly exactly cancel out.
A small error in either of them produces a significant error in the total energy transfer rate.
In the electron equation this transfer rate is given by the difference in the synchrotron cooling and heating rates.
To maintain the energy balance between the two equations, we need to ensure that in our numerical scheme these rates
are seen identically by both equations.

In discretized form, the synchrotron processes for electrons/positrons are described by equations (\ref{eq42})--(\ref{eq43}),  
with $n=\npoel$, $A= A_{\rm e,syn}$ and $B=B_{\rm e,syn}$.
To obtain the total energy gain we have to multiply equation~ (\ref{eq42}) by $\gamma_i \,\gridp$,
sum over $i$ and sum the corresponding terms in the electron and positron equations. Assuming vanishing boundary currents, we have
\begin{equation}\label{eq52}
\frac{\Delta E_{\rm e}}{\Delta t_{k}\: \Delta V} = 
\sum_{i=1}^{i_{m} - 1} \Delta\gamma_{i+1/2}  \left[  
A_{i+1/2} \; \distr_{{\rm e},i+1/2}  -  B_{i+1/2}  \frac{\distr_{{\rm e},i+1} 
- \distr_{{\rm e},i} }{\gridp} \right],			
\end{equation}
where $\Delta\gamma_{i+1/2} \equiv \gamma_{i+1} - \gamma_{i}$
and we have omitted the time index $k+1/2$ for brevity. 
The exchange rate as seen by the photon equation can be evaluated by writing the integrals in emissivity and absorptivity
expressions (\ref{eq33}) and (\ref{eq35a}) as sums over the grid, multiplying equation~(\ref{eq901}) by $x_l \,\gridx$ and 
summing over $l$: 
\beq\label{eq53}
& & \frac{\Delta E_{\rm ph}}{\Delta t_{k}\: \Delta V}
= \sum_{l=1}^{l_m} \left[- c \: x_l \: \alpha_l  \: \distr_{{\rm ph},\: l}  + 
x_l \: \epsilon_{\, l}  \right] \gridx
\\
& =&    \frac{\lambdac^3}{8\pi}  \sum_{l=1}^{l_m} \left[ \frac{\distr_{{\rm ph},\: l} }{x_l} \: \gridx 
\sum_{i=1}^{i_m} \frac{\gamma_i P_{l,i}}{p^2_i}  \gridp
\left( -3 \distr_{{\rm e},i}  \: 
+ \frac{\distr_{{\rm e},i+1}  - \distr_{{\rm e},i} } {\gridp} \right) \right]
+   \sum_{l=1}^{l_m} \left[ x_l \: \gridx \sum_{i=1}^{i_m} \distr_{{\rm e},i}  \: P_{l,i} \: \gridp \right] , 	\nonumber		
\eeq
where $P_{l,i}=P(x_l,p_i)$. Changing the order of summation, identifying the sum over the photon distribution as the discretized version of the definition $H(p)$, and noticing that $\sum_l P_{l,i} x_l \gridx$ gives the electron cooling rate $-\dot{\gamma}_{{\rm s},i}$, we get: 
\begin{equation}\label{eq55}
\frac{\Delta E_{\rm ph}}{\Delta t_{k}\: \Delta V} 
 =  \sum_{i=1}^{i_m} \gridp \left[ -\left( \dot{\gamma}_{{\rm s},i} 
+ \frac{3\gamma_i H_{i} }{p^2_i} \right) \distr_{{\rm e},\: i} 
+ \frac{\gamma_i H_{i} }{p^2_i} \frac{\distr_{{\rm e},\: i+1}  - \distr_{{\rm e},\: i} }{\gridp} \right].   
\end{equation}
To make equations (\ref{eq52}) and (\ref{eq55}) identical (except for the sign) we have to make subtle changes in the definition of
coefficients and the way integrals are numerically calculated. In equation~(\ref{eq55}) we have to
define the coefficients in between the electron momentum gridpoints, at $i+1/2$, 
substitute the electron distribution $\distr_{{\rm e},\: i}$ by $\distr_{{\rm e},\: i+1/2}$ (except in the derivative term), where the latter is calculated using the same Chang \& Cooper coefficients $\delta_i$ as in the electron equation, and sum up to $i=i_m-1$ instead of $i_m$.
This amounts to defining the emission and absorption coefficients as
\begin{equation}\label{eq56}
 \epsilon_l  =
\sum_{i=1}^{i_m - 1} P_{l,i+1/2} \: \distr_{{\rm e},\: i+1/2}  \: \gridp, 
 \quad 
\alpha_l  = 
 \frac{\lambdac^3}{8\pi}  \frac{1}{x_l^2} 
\sum_{i=1}^{i_m-1} \frac{\gamma_{i+1/2} \: P_{l,i+1/2}}{p_{i+1/2}^2}
\left[ 3 \distr_{{\rm e},i+1/2}  - \frac{\distr_{{\rm e},i+1}  - \distr_{{\rm e},i} }{\gridp} \right] \:
\gridp.								 
\end{equation}
Also, the coefficients $A$ and $B$ entering 
the momentum space flux (\ref{eq43}) and thus also
the electron energy exchange rate (\ref{eq52}) should be written as
\begin{equation}
  A_{i+1/2}  = \frac{\gridp}{\Delta\gamma_{i+1/2}}
\left(\dotgammas + 3 \frac{\gamma}{p^2} H \right)_{i+1/2}   \quad \mbox{and} \quad
B_{i+1/2}  = \frac{\gridp}{\Delta\gamma_{i+1/2}} \left(\frac{\gamma}{p^2} H \right)_{i+1/2} 
 ,
\end{equation}
which become identical to (\ref{eq:AB_elsyn}) in the limit $\gridp \rightarrow 0$
and ensure that the energy exchange rates as seen by the electron and photon equations are the same.

The only discrepancy left  is that  we cannot use the same $\distr_{\rm ph}^{k+1/2}$ and $\distr_{\rm e}^{k+1/2}$ in both equations. This is because each of them contains a function $\distr^{k+1}$, which, in the equation that we evolve before, is not known for the other type of particle. The solution to this, at least in the average sense, is to regard the time-grids for each equation as shifted by a half timestep. Then $\distr^{k+1}$ obtained from one equation can be used as $\distr^{k+1/2}$ in the other and vice versa.

\subsection{Coulomb collisions}

Coulomb scattering only redistributes the energy between different parts of the lepton population.
It is easy to see that the total energy is conserved in the sum of two equations (\ref{Coulnum:eq1}) for electrons and positrons, provided that $a(\gamma,\gamma_1)$ is antisymmetric, the latter simply reflects the energy conservation in two-body interactions. Similarly to synchrotron, our numerical treatment has to ensure that the conservation is exact, otherwise unphysical runaways can occur near the equilibrium.

The flux in momentum space in equation (\ref{eq42}) for Coulomb scattering is given by equation (\ref{eq43})
with coefficients expressed as (see eq. [\ref{Coul:AB}])
\begin{equation} \label{Coulnum:eq3}
A_{i+1/2} =   \left(\frac{\dot{\gamma} \gamma}{p^2} \right)_{i+1/2}
- \frac{1}{2\Delta\gamma_{i+1/2}} \left[ \left( \frac{\gamma \, D}{p^2} \right)_{i+1}
- \left( \frac{\gamma \, D}{p^2} \right)_{i} \right] , 
\qquad B_{i+1/2} =  \frac{1}{2} \left(\frac{\gamma^2 D}{p^4}\right)_{i+1/2}.		
\end{equation}
The total energy exchange rate is identical to equation (\ref{eq52}) for synchrotron.

Let us now look separately at terms containing $\dot{\gamma}$ and $D$. 
For $\dot{\gamma}$ we have 
\beq \label{Coulnum:eq5}
\left. \frac{\Delta E_{\rm e}}{\Delta t_{k}\: \Delta V} \right|_{\dot{\gamma}}
=  \sum_{i=1}^{i_{m} - 1} \Delta\gamma_{i+1/2} \left(\frac{\dot{\gamma} \, \gamma}{p^2} \right)_{i+1/2} \;
\distr_{{\rm e},i+1/2}.										
\eeq
It is now easy to see that this quantity can be made to vanish if we write
\begin{equation} \label{Coulnum:eq6}
\dot{\gamma}_{i+1/2} = \sum_{l=1}^{i_{m} - 1} a(\gamma_{i+1/2},\gamma_{l+1/2}) \; \distr_{{\rm e},l+1/2} \, \gridp
\quad \mbox{and} \quad
\left(\frac{\dot{\gamma} \, \gamma}{p^2} \right)_{i+1/2} \rightarrow  \dot{\gamma}_{i+1/2} \frac{\gridp}{\Delta\gamma_{i+1/2}},
\end{equation}
provided that $a$ is antisymmetric. The terms containing $D(\gamma)$ in the energy exchange rate are
\begin{equation} \label{Coulnum:eq7}
\left. \frac{\Delta E_{\rm e}}{\Delta t_{k}\: \Delta V} \right|_{D}
=  -\frac{1}{2} \sum_{i=1}^{i_{m} - 1}
\left\{ \left[ \left( \frac{\gamma \, D}{p^2} \right)_{i+1}
- \left( \frac{\gamma \, D}{p^2} \right)_{i} \right]  \distr_{{\rm e},i+1/2}  
+
\left(\frac{\gamma D}{p^2}\right)_{i+1/2}
\left( \distr_{{\rm e},i+1} - \distr_{{\rm e},i} \right)
\right\},										
\end{equation}
where we have redefined the coefficient $B$ as
\begin{equation} \label{Coulnum:eq8}
B_{i+1/2} \rightarrow \frac{1}{2} \frac{\gridp}{\Delta\gamma_{i+1/2}} 
\left(\frac{\gamma D}{p^2}\right)_{i+1/2}. 	
\end{equation}
One can see that equation (\ref{Coulnum:eq7}) has the form of an integral over a full differential and, as such, should vanish provided that $D=0$ at the boundaries. To ensure this numerically for any electron distribution we write explicitly 
$\distr_{{\rm e},i+1/2} = (1-\delta_i) \distr_{{\rm e},i+1} + \delta_i \distr_{{\rm e},i}$
and demand that the coefficient in front of $\distr_{{\rm e},i}$ in equation (\ref{Coulnum:eq7}) is equal to zero for every $i$. Rearranging terms, we get
\beq \label{Coulnum:eq9}
\left. \frac{\Delta E_{\rm e}}{\Delta t_{k}\: \Delta V} \right|_{D} &=&
-\frac{1}{2} \sum_{i=2}^{i_{m} - 1} 
\distr_{{\rm e},i} \left\{ \delta_i 
\left[ \left( \frac{\gamma \, D}{p^2} \right)_{i+1}
- \left( \frac{\gamma \, D}{p^2} \right)_{i} \right]
+ (1-\delta_{i-1})
\left[ \left( \frac{\gamma \, D}{p^2} \right)_{i}
- \left( \frac{\gamma \, D}{p^2} \right)_{i-1} \right]
\right.	\nonumber \\
&-& \left. \left(\frac{\gamma D}{p^2}\right)_{i+1/2} + \left(\frac{\gamma D}{p^2}\right)_{i-1/2}
\right\}
+ \distr_{{\rm e},1} S^{-}
+ \distr_{{\rm e},i_{m}} S^{+}.							
\eeq
The expression in the curly brackets is identically zero if we set
\begin{equation}  \label{Coulnum:eq12}
\left(\frac{\gamma D}{p^2}\right)_{i+1/2} = \delta_i \left(\frac{\gamma D}{p^2}\right)_{i+1}
+ (1-\delta_i) \left(\frac{\gamma D}{p^2}\right)_{i},						
\end{equation}
while the boundary terms $S^{-}$ and $S^{+}$ vanish if
\begin{equation} 	\label{Coulnum:eq13}
\left(\frac{\gamma D}{p^2}\right)_{1} = 0 \quad \mbox{and} \quad \left(\frac{\gamma D}{p^2}\right)_{i_m} = 0.
\end{equation}
Using expressions (\ref{Coulnum:eq6}) in the first term in coefficient $A$ 
and equations (\ref{Coulnum:eq12}) and (\ref{Coulnum:eq13}) in the definition (\ref{Coulnum:eq8}),
we ensure precise energy conservation in the numerical scheme. 
%

\begin{figure*}
\begin{center}\leavevmode\epsfxsize=17.2cm \epsfbox{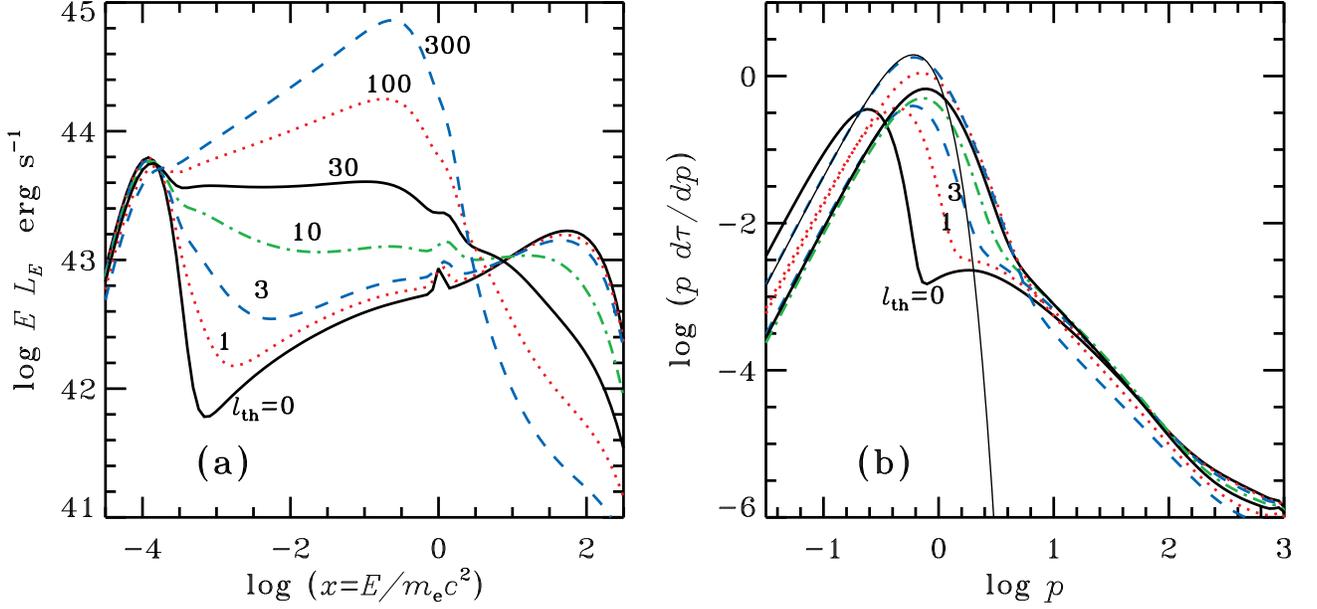}\end{center}
\caption{Equilibrium ({\it a}) photon spectra and ({\it b}) electron distributions (Thomson optical depth per $\ln p$, i.e. $\nel(p)\sigmat R$) for various  stochastic heating compactnesses $l_{\rm th}$ as labeled. The size of the emission region is $R = 10^{14}$ cm, the soft input radiation has a compactness $l_{\rm s} = 10$ and a blackbody temperature $T_{\rm BB} = 15$ eV, the injection compactness is $l_{\rm nth} = 10$.  The thin solid line on the right panel
shows a Maxwellian fit of temperature $T_{\rm e} = 53$ keV. Compare to fig.~1 in \citet{coppi99}. } 
\label{fig1}
\end{figure*}

\section{Numerical results}

Our careful treatment of the micro-physical processes makes the code applicable over a wide range of parameter regimes. The current version covers 15 orders of magnitude in photon energy (from $10^{-5}$ to  $10^{10}$ eV) and 8 orders of magnitude in electron momentum, while there is no fundamental difficulty in extending this range further, e.g. to TeV energies for application to blazars. Energy conservation is achieved to within 1\% in the majority of cases. All the rates and cross-sections of different processes have been precalculated once and for all and are read into memory as the code initializes. A typical simulation for 200 grid points in photon energy and electron momentum on a 3 GHz PC running Linux takes between a few minutes and half an hour. 

In order to test the performance of our code in different parameter regimes, we have chosen three setups from earlier works and run the code with similar parameters for comparison. 

\subsection{Non-thermal pair model}

As a first test we  compare our code to the well-known pair plasma code \eqpair\ by \citet{coppi92,coppi99}. \eqpair\ also considers an uniform emission region into which high-energy electrons/pairs are injected, mimicking an unspecified acceleration mechanism. Some low-energy photons are also injected, emulating a source of external soft radiation (e.g. accretion disk). The high-energy pairs cool by Compton scattering 
and Coulomb energy exchange with colder thermal pairs.
The Compton upscattered photons can produce electron-positron pairs which then upscatter more photons etc., initiating a pair cascade. Once the pairs cool down to low enough energies, the timescale of the systematic energy losses becomes longer than that of diffusive processes, leading to relaxation into a low-energy thermal distribution.
In \eqpair, Coulomb collisions between particles are assumed to be the thermalizing mechanism.
However, the thermalization process is not treated entirely consistently in this code in a sense that there exists only one thermal bin into which particles are put once they have cooled below a certain threshold  energy, chosen to be $\gamma = 1.3$. The electron temperature associated with this thermal bin is nevertheless calculated self-consistently from energetic considerations. Furthermore, the code does not consider thermalization by synchrotron self-absorption, which can be an efficient mechanism if the medium is magnetized \citep[][ GHS98]{GGS88}.

The setup of this test run is similar to what was used for fig.~1 in \citet{coppi99}.  We switched off synchrotron processes in our code and left other  processes. We inject a Gaussian distribution of pairs centered at $\gamma_{\rm inj} = 10^3$ and a low-energy blackbody distribution of photons. In addition, there is a background electron plasma present with optical depth $\tau_{\rm p} = 0.1$. There is no escape term for pairs, meaning that all injected pairs eventually annihilate transferring their energy to the radiation field. The power injected as non-thermal pairs is parametrized by compactness
\begin{equation}
l_{\rm nth} = \frac{\sigmat}{\me c^3} \frac{L_{\rm nth}}{R},									\label{eq113e}
\end{equation}
where $L_{\rm nth}$ is the injected luminosity  (including rest mass) and $R$ is the linear dimension of the emission region.
Similarly, we define the compactness of the injected soft radiation as
\begin{equation}
l_{\rm s} = \frac{\sigmat}{\me c^3} \frac{L_{\rm s}}{R},									\label{eq113f}
\end{equation}
where $L_{\rm s}$ is the relevant luminosity. To mimic acceleration with less than 100\% efficiency,
additional power is supplied to low-energy electrons in the form of continuous heating,
parametrized by $l_{\rm th}$.
In \citet{coppi99} this energy was just given to the thermal bin, but since we do not have such bin in our code, we need to explicitly specify the form of this heating. This is done by stochastic acceleration prescription of the form 
\begin{equation}
\left.\frac{D \noccpm(p)}{Dt}\right|_{\rm{stoch.}} = \frac{1}{p^2} \frac{\partial}{\partial p}
\left[ p^2 \Dacc(p) \frac{\partial \noccpm (p)}{\partial p}
\right].															\label{eq113g}
\end{equation}
The momentum diffusion coefficient
is assumed to take the form characteristic of
stochastic acceleration by resonant interactions with 
plasma waves \citep{DML96},		
$\Dacc(p) \propto p^q$.
We have chosen $q = 2$ in our calculations. 
The mean energy gain rate of a particle resulting from equation~(\ref{eq113g}) is
\begin{equation}
\left. \left\langle \frac{\rmd \gamma}{\rmd t} \right\rangle \right|_{\rm stoch.} = 
\frac{1}{p^2} \frac{\partial}{\partial p} \left[ \beta p^2 \Dacc(p) \right],		\label{eq113g1}
\end{equation}
where $\beta=p/\gamma$ is the particle speed. We can see that for a power-law diffusion coefficient the gain rate is proportional to $p^{q-1}$ in the relativistic regime, while in the nonrelativistic regime it is proportional to $p^q$.
Our choice $q=2$ means that at high energies Compton losses always overcome gains by stochastic acceleration, the main effect of the latter process is therefore the heating of low-energy pairs.

The differential term given by equation~(\ref{eq113g}) is included in the Chang \& Cooper scheme on the same grounds with other continuous terms.
Therefore before discretization it has to be written in the form compatible with equations~(\ref{eq42}) and (\ref{eq43}):
\begin{equation}
\left.\frac{D \npoel (p)}{Dt}\right|_{\rm{stoch.}} = -\frac{\partial}{\partial\ln{p}}
\left\{ \Dacc(p) \frac{1}{p^2} \left[ 3 \npoel(p) - \frac{\partial \npoel (p)}{\partial\ln{p}}
\right] \right\}.															\label{eq113h}
\end{equation}

The results of the test are shown in Fig.~\ref{fig1}.
Varying the amount of stochastic heating ($l_{\rm th}$)
keeping all other parameters constant, we see that we can well reproduce the behavior of the spectrum in fig.~1 in \citet{coppi99}. Just as expected by \citet{coppi99}, the equilibrium electron distribution is hybrid: Maxwellian at low energies with a nonthermal high-energy tail. Note that we get such shape even if we switch off Coulomb scattering.  
The thermal-looking distribution is produced by the stochastic heating itself,
which gives a Maxwellian slope at low energies irrespective of the shape of $\Dacc(p)$, while the location of the peak of the distribution is determined by the balance between heating and Compton cooling. 
The behavior of the spectrum in response to varying the power of stochastic heating seen in Fig.~\ref{fig1}{\it a} was analyzed in detail by \citet{coppi99}, we are not going to repeat it here.

\subsection{Thermalization by synchrotron self-absorption}

\begin{figure*}
 \begin{center}\leavevmode\epsfxsize=17.2cm \epsfbox{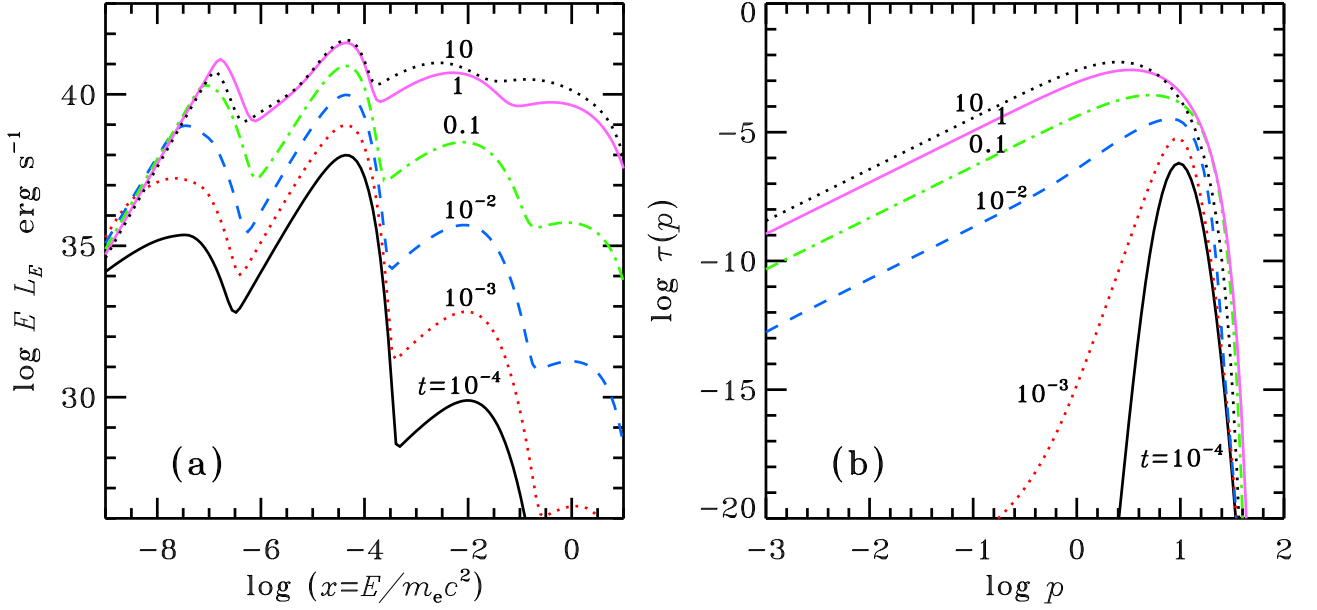}\end{center}
\caption{Evolving ({\it a})  photon spectra and ({\it b}) electron  distributions ($\tau(p)=\sigmat R \nel(p)/p$)  for Gaussian electron injection under action of Compton and synchrotron processes at different times (in $R/c$ units)  as labeled. 
The source size is $R = 10^{13}$ cm, the magnetic compactness is $l_{\rm B} = 10$
and the injection compactness $l_{\rm nth} = 1$. Compare to fig.~1 in GHS98.}
\label{fig2}
\end{figure*}

\begin{figure*}
\begin{center}\leavevmode\epsfxsize=17.2cm \epsfbox{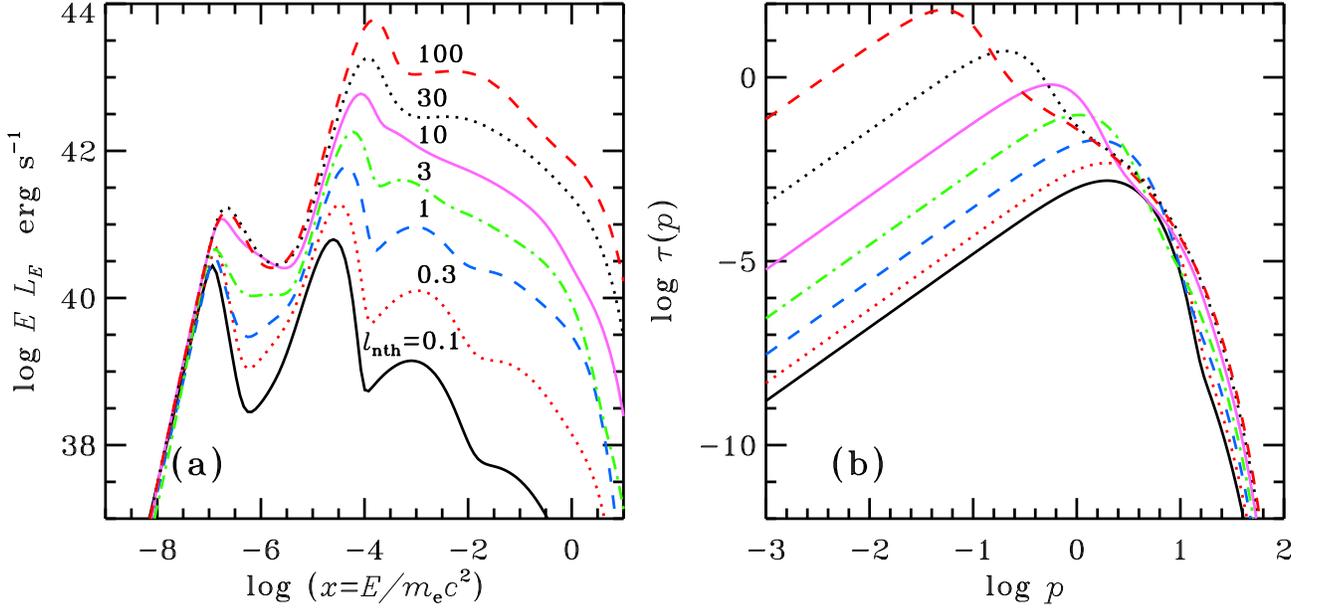}\end{center}
\caption{Equilibrium ({\it a})  photon spectra and ({\it b}) electron  distributions for injection (\ref{eq:qe})  for 
various injection compactnesses $l_{\rm nth}$ as labeled. Parameters: $R = 10^{13}$ cm, $l_{\rm B} = 30$. Compare
 to fig.~2 in GHS98.}		
\label{fig3}
 \end{figure*}

For the second test, we compared our results with these of GHS98. They studied electron thermalization by synchrotron self-absorption in the presence of Compton cooling. The electron cooling, heating and diffusion due to the synchrotron were described by equation~(\ref{eq36}) (without the last term), while Compton scattering was assumed to take place in the Thomson regime and contribute only to systematic cooling. Furthermore, the treatment was not fully self-consistent since only the electron equation was actually solved. While the equilibrium synchrotron spectrum was self-consistently calculated at each timestep from the formal solution of the radiative transfer equation, the Comptonized spectrum was not. Thus only the synchrotron spectrum entered the electron heating rate by self-absorption, while the radiation energy density needed to account for Compton cooling was estimated from energetic considerations.			 

We ran our code with the same parameters used to obtain the results in figs. 1 and 2 in GHS98. The pair production/annihilation and Coulomb scattering have been switched off for this test. High-energy electrons are injected into the emission region, with the total power (including rest mass) parametrized by the injection compactness $l_{\rm nth}$. The magnetic compactness is defined by
\begin{equation}
l_{\rm B} = \frac{\sigmat}{\me c^2} R U_{\rm B},
\end{equation}
where $U_{\rm B}$ is the magnetic energy density. In addition there is an external source of soft blackbody photons assumed to arise from reprocessing half of the hard radiation by cold matter in the vicinity of the emission region. The electron escape timescale is fixed at $t_{\rm esc} = R/c$.

In the first case the injected electrons have a Gaussian distribution peaking at $\gamma=10$. The evolution of this distribution is followed in time as it cools and thermalizes by Compton and synchrotron processes. We can see that our results shown in Fig.~\ref{fig2} are almost identical to those presented in fig.~1 in GHS98. However, we would like to stress that we also compute self-consistently the photon spectrum. We see the partially self-absorbed synchrotron bump at small energies,
then the blackbody photons and two Compton scattering orders at higher energies.

In the second case we calculated the steady-state particle distributions for different injection compactnesses. The injected electron distribution (per unit $\ln p$) is 
\begin{equation} \label{eq:qe}
Q_{\rm e} = Q_0 \, \frac{p^3}{\gamma^2} \exp{\left(-\frac{\gamma}{\gamma_{\rm c}} \right)},
\end{equation}
where $\gamma_{\rm c} = 3.33$.
The resulting equilibrium electron distributions plotted in Fig.~\ref{fig3}{\it b} are again very similar to the ones obtained by GHS98 in their fig.~2. The corresponding radiation spectra shown in Fig.~\ref{fig3}{\it a} are computed self-consistently and simultaneously with the electron distribution (while the spectra in fig.~4 of GHS98 are calculated a posteriori, i.e. after the equilibrium electron distribution has been determined). 		 
As discussed in GHS98, if the source is strongly magnetically dominated, the equilibrium distribution is almost purely Maxwellian. When the injection compactness increases, Compton losses become non-negligible  and the electrons cool down to lower energies before they have time to thermalize. Notice that at the highest compactness ($l_{\rm nth} = 100$) the temperature of the Maxwellian part of the distribution inferred from Fig.~\ref{fig3}{\it b} deviates appreciably from the one obtained by GHS98. This is caused by the fact that at high compactness a significant fraction of the soft radiation is Compton upscattered to energies comparable to the energies of the Maxwellian electrons. These photons are therefore not effective in cooling the electrons any further. However, in GHS98 Compton cooling is accounted for through a term proportional to the radiation energy density, which includes all photons, and therefore overestimates the cooling rate.  Overall, the simple prescription for Compton cooling without actually solving the photon equation appears to work well in the parameter regimes considered here.

\begin{figure*}
\begin{center}\leavevmode\epsfxsize=17.2cm \epsfbox{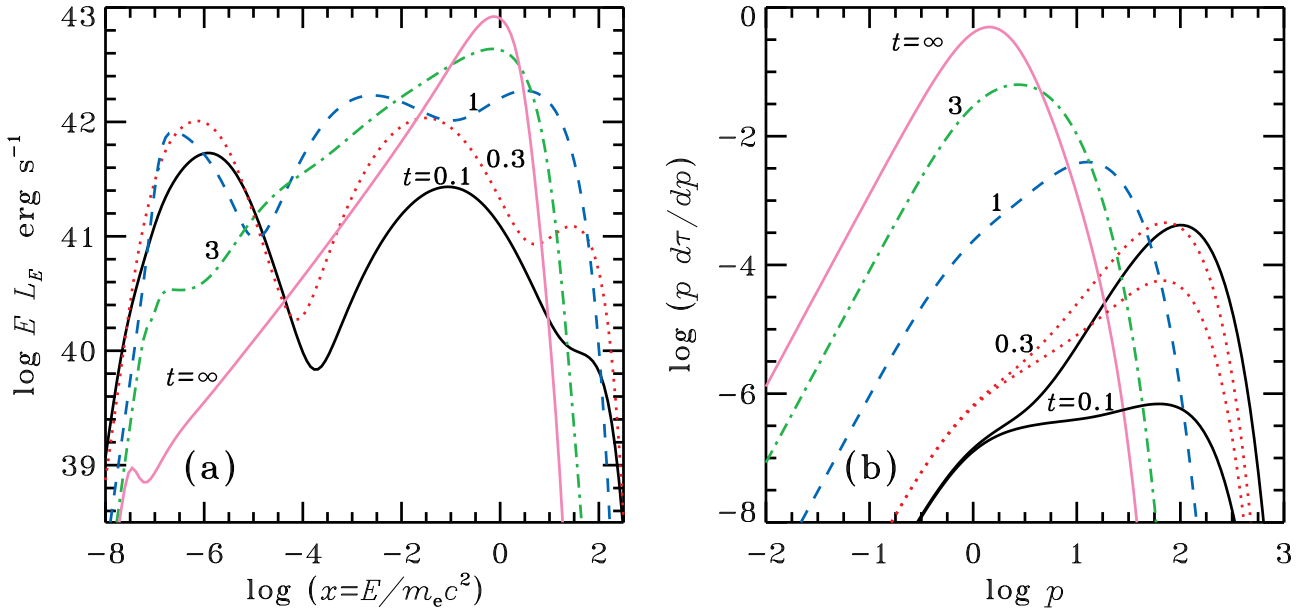}\end{center}
\caption{Evolving ({\it a}) photon spectra and ({\it b}) Thomson optical depth per $\ln p$ for stochastically heated pairs
at different times (in units $R/c$) as labeled. Parameters: the source size $R = 10^{13}$ cm, the magnetic compactness $l_{\rm B} = 0.3$, the stochastic heating compactness $l_{\rm th} = 30$, the initial Thomson optical depth of electrons is $\tau_0 = 6\times 10^{-4}$.  For $t = 0.1, 0.3$ we also plot positrons, at later times
only the electrons as their opacities are nearly identical. 
Compare to fig.~2 in \citet{SP04}.}
\label{fig4}
\end{figure*}

\subsection{Gamma-ray bursts from stochastically heated pairs}

Finally, we compare our code to the Large Particle Monte Carlo code by \citet{SBS95}, with all the processes operating now. The setup is similar to the one used in \citet{SP04} for simulating the spectral evolution of gamma-ray bursts. They consider an initially optically thin distribution of electrons in a cylinder-shaped emission region. Arguing that impulsive first-order Fermi acceleration would result in cooling spectra that are too soft to be consistent with observations, energy is instead supplied to the electrons continuously, mimicking dissipation by plasma instabilities behind the shock front. As electrons are heated to relativistic energies in the prescribed background magnetic field, they emit synchrotron radiation, providing seed photons for Compton upscattering. The high-energy upscattered photons then initiate pair-production.

In our simulation we consider a spherical region permeated by magnetic field and start by heating a cold electron distribution (with initial Thomson optical depth $\tau_0 = 6\times 10^{-4}$) according to the stochastic acceleration prescription (\ref{eq113h}). No pair escape is allowed. The results of simulations are shown in Fig.~\ref{fig4} and can be compared to a similar fig.~2 in \citet{SP04}. In both cases the electrons are rapidly heated to about $\gamma \sim 100$, as determined by the balance between stochastic heating and synchrotron cooling. As the photon field builds up, additional cooling by Compton scattering causes the electron 'temperature' to start dropping. After about 1/3 of the light crossing time, the number of photons upscattered to the MeV range becomes large enough to start significant pair-production. With the increasing pair density (at $t=1$, opacity has grown by a factor of 20) the available energy per particle decreases, causing a further drop in the temperature of the now almost pure pair plasma. After about ten light-crossing times the Thomson opacity is $\tau_{\rm T}=1.3$ and the pair density reaches the value where the pair annihilation and creation rates are balanced and a steady state is attained.

The spectral behavior seen in Fig.~\ref{fig4}{\it a} is similar to what was obtained by \citet{SP04}. The synchrotron peak rises first, being initially in the optically thin regime and thus following the evolution of the peak of the electron distribution according to $x \propto \gamma^2$. The first Compton scattering order lags slightly behind synchrotron, while the second scattering order is initially in Klein-Nishina regime and thus hardly visible at all. As the electron temperature drops and the peak of the first scattering order evolves to lower energies, the second order shifts to the Thomson regime and becomes comparable to and eventually dominant over the first order. At the same time the decreasing temperature and increasing pair opacity causes the synchrotron emission to switch to optically thick regime and the synchrotron luminosity to drop dramatically. The plasma becomes photon starved and the Comptonized spectrum hardens.

\section{Conclusions}

We have developed a new computer code for simulations of the radiative processes in magnetized rarefied plasmas encountered in the vicinities of accreting black holes and relativistic jets in active galaxies and gamma-ray bursts.  We take into account Compton scattering, pair production and annihilation,  synchrotron processes and Coulomb scattering without any limitations on the energies of the photons and electrons/positrons. We solve coupled  integro-differential kinetic equations describing time evolution of the photon and electron/positron  distributions simultaneously. The equations contain both integral and up to second order differential terms.  
The Fokker-Planck  differential terms are substituted when necessary instead of the  integral terms with coefficients computed exactly from the moments of the integral equation. This allows us to study thermalization of the lepton distribution by Compton and Coulomb scattering and synchrotron self-absorption. Processes involving bremsstrahlung can be easily added to the code, while for the conditions considered in the paper they are not important. 

The presented technique guarantees energy (and particle, when relevant) conservation with high accuracy which is especially important when dealing with strongly self-absorbed synchrotron radiation. The implementation of the Chang and Cooper scheme gives the correct shape of the particle distribution at low energies. The area of application of the code is enormous as it can deal with photons and leptons covering many orders of magnitude in momentum space, with no potential difficulty of extending it to even lower/higher energies. We present a number of test runs, where we consider problems previously solved by other methods. We compute non-thermal pair cascades, and study lepton thermalization by synchrotron self-absorption, as well as model the emission from the stochastically heated pairs that might have a relation to the prompt emission of gamma-ray bursts. We find a good agreement in the parameter space where comparison is feasible while the differences can be explained by our improved treatment of microphysics.

\acknowledgments

We are grateful to Dmitrij Nagirner and Paolo Coppi for a number of useful discussions and suggestions. This work was supported by the CIMO grant TM-06-4630, the Magnus Ehrnrooth Foundation, and the  Academy of Finland grants 110792 and 122055.

\appendix

\section{Relation between the Compton redistribution functions for photons and electrons}
\label{app:relation}

The redistribution functions defined by equations~(\ref{eq12}) and (\ref{eq20}) can be written as
\begin{equation}\label{eq114}
\overRph(x,x_1,\gamma_1)  =   \frac{1}{4\pi^2}\: p_1
\int p \: \rmd\gamma \: \rmd^2\Omega \: \delta(\gamma_1 + x_1 - \gamma - x)
\int \rmd^2 \Omega_1 \: \rmd^2 \omega_1 \: F \:
\delta(\vecp_1 + \vecx_1 - \vecp - \vecx)	,	
\end{equation}
\begin{equation} \label{eq115}
\overRe(\gamma,\gamma_1,x_1)  =  \frac{1}{4\pi^2} \: x_1
\int x \: \rmd  x \: \rmd^2\omega \: \delta(\gamma_1 + x_1 - \gamma - x)
\int \rmd^2\Omega_1
\: \rmd^2\omega_1 \: F \: \delta(\vecp_1 + \vecx_1 - \vecp - \vecx) .			
\end{equation}
We see that the inner integrals are identical in both expressions. Because of rotational symmetry, the only angle left in the calculation after performing the integrals over $\rmd^2\Omega_1 \rmd^2\omega_1$ is the angle between the momenta of outgoing particles. Therefore we can write $\rmd^2\Omega = \rmd^2 \omega = 2\pi \rmd\zeta$, where $\zeta = \vOmega\cdot\vomega$.
We also see that $\rmd\gamma \: \delta(\gamma_1 + x_1 - \gamma - x) = \rmd x\: \delta(\gamma_1 + x_1 - \gamma - x)$,
so we find from equations~(\ref{eq114}) and (\ref{eq115}) that the redistribution functions are related as
\begin{equation}\label{eq116}
pp_1 \overRe(\gamma,\gamma_1,x_1) = xx_1 \overRph(x,x_1,\gamma_1),			
\end{equation}
where one of the energies/momenta has to be replaced from the condition $x + \gamma = x_1 + \gamma_1$.

\section{Compton redistribution function}
\label{app:crf}

The isotropic Compton redistribution function defined in equation~(\ref{eq12}) can be written as an integral over the scattering angle (NP94) 
\begin{equation}\label{eq117}
\overRph(x,x_1,\gamma_1) = \left. \int_{\mu_m}^{\mu_{+}} R(x,x_1,\gamma_1,\mu) \ \rmd\mu 
= T(x,x_1,\gamma_1,\mu) \ \right|_{\,\mu_{\rm m}}^{\,\mu_{+}}. 
\end{equation}
The limits of integration are given by
\begin{equation}\label{eq118}
\mu_{\rm m} = 
\begin{cases} 
-1 &\text{if $|x-x_1| \ge 2xx_1$,} \\
-1 &\text{if $|x-x_1| \le 2xx_1$ and $\gamma_1 \ge \gamma_{\star}(x,x_1,-1)$,} \\
\mu_{-} &\text{if $|x-x_1| \le 2xx_1$ and $\gamma_{\rm m}  \le \gamma_1 \le \gamma_{\star}(x,x_1,-1)$,} 
\end{cases}
\quad \mu_{-} = 1 - \frac{D_{\rm m}}{xx_1}, \quad \mu_{+} = 1 - \frac{(x-x_1)^2}{D_{\rm m} xx_1},
\end{equation}
where
\begin{equation}\label{eq120}
D_{\rm m} = p_1^2 + \gamma_1 (x_1 - x) + p_1 \sqrt{(\gamma_1 + x_1 - x)^2 - 1}	 , 
\qquad \gamma_{\rm m}  = 1+(x-x_1+ |x-x_1|)/2. 
\end{equation}
The quantity $\gamma_{\star}(x,x_1,-1)$ is the minimum electron energy needed to scatter 
a photon backwards  (i.e. $\mu = -1$) from $x_1$ to $x$:
\begin{equation}\label{eq121}
\gamma_{\star}(x,x_1,-1) = [ x - x_1 + (x+x_1)\sqrt{1+1/xx_1} ]/2.
\end{equation}

The angle-dependent redistribution function $R(x,x_1,\gamma_1,\mu)$ was first derived by \citet{AA81}, see also \citet{PKB86} and \citet{NP93}. The angle-averaged function was obtained by \citet{Jones68}, but the presented expressions are  very cumbersome and the loss of accuracy  occurs for small photon energies and large electron energies. An alternative function given by \citet{Bri84} and NP94  does not suffer from these  problems.  We use here the latter expressions. The primitive function $T(x,x_1,\gamma_1,\mu)$ can be expressed through functions of one argument as
\begin{equation}
\label{eq122}
T(x,x_1,\gamma_1,\mu) =  -\frac{2}{xx_1} Q + \sqrt{\frac{w}{2}}
\left\{
\frac{4}{xx_1} H_0 + w \left( 1 + \frac{1}{xx_1} \right) H_1 
+\frac{H}{A(h_{-}) A(h_{+})} \left[ w + \frac{1}{2 x^2 x_1^2} \left( 2\frac{H^2}{w} - (x-x_1)^2 \right) \right]
\right\},																
\end{equation}
where $w = 1-\mu$ and $Q = \sqrt{(x-x_1)^2 + 2xx_1 w}$.
The functions $H$ are given by the differences
\begin{equation}\label{eq123}
H = A(h_{-}) - A(h_{+}), \quad H_n = A_n(h_{-}) - A_n(h_{+}),	
\end{equation}
where  
\begin{equation} \label{eq128}
A(h) = \sqrt{1+h}	,\qquad h_{+} = [ (\gamma_1 + x_1)^2 - 1 ] \, w/2,\qquad h_{-} = [ (\gamma_1 - x)^2 - 1 ] \, w/2 . 
\end{equation}
The zeroth function $A_0$ is
\begin{equation}\label{eq126}
A_0(h) = 
\begin{cases} 
\ln(\sqrt{h} + \sqrt{1+h})/\sqrt{h}  &\text{if $h \ge 0$,} \\
\arcsin(\sqrt{-h})/\sqrt{-h}             &\text{if $h \le 0$,}
\end{cases}
\end{equation}
while the others can be expressed through its derivatives as
\begin{equation}\label{eq127}
A_n(h) = (-2)^n \frac{|2n-1|}{(2n-1)!!} A_0^{(n)}(h) ,										
\end{equation}
and can be 
computed by the recurrent relation
\begin{equation} \label{eq:AnAn}
A_{n+1} (h)=\frac{1}{h}\left[ \frac{2n+1}{|2n-1|}A_n (h)-\frac{1}{A^{2n+1}(h)} \right],
\end{equation}
or  for $|h|\leq 1$ via series 
\begin{equation} \label{eq:Anser}
A_n(h)=\frac{|2n-1|}{(2n-1)!!}\sum_{k=0}^{\infty}\,
\frac{(2n+2k-1)!!}{(2k)!!}\frac{(-h)^k}{2n+2k+1}.
\end{equation}
Direct computations using (\ref{eq123}) lead to numerical cancellations at small photon energies
$x, x_1 \ll 1$. NP94 describe in details how they should be dealt with.

\section{Moments of the Compton redistribution function}
\label{app:moments}

The moments of the photon redistribution function given by equation~(\ref{eq113a}) can be written explicitly
using equation~(\ref{eq3}) as
\begin{equation} \label{eq129}
\overline{x_1^i}   \, \overline{s}_0(x) 
= \frac{3}{16\pi} \frac{2}{\lambdac^3\Nel} \frac{1}{x}
\int \frac{\rmd^3 p}{\gamma} \frac{\rmd ^3 p_1}{\gamma_1} \frac{\rmd ^3 x_1}{x_1} \, x_1^i \, \nocce(\vecp) \: F
\: \delta^4,												
\end{equation}
where we have denoted $\delta^4 = \delta(\fourp_1 + \fourx_1 - \fourp - \fourx)$ for brevity.
We now define (NP94)
\begin{equation} \label{eq130}
\langle x_1^i \rangle \, s_0(\xi ) = \frac{3}{16\pi} \frac{1}{\xi } \int \frac{\rmd^3 x_1}{x_1} \frac{\rmd ^3 p_1}{\gamma_1} \, x_1^i
\: F \: \delta^4											
\end{equation}
and
\begin{equation}\label{eq131}
\Psi_i(x,\gamma) = \frac{1}{4\pi\gamma x^{i+1}} \int \rmd^2\Omega\ \, \xi\, \langle x_1^i\rangle  \, s_0(\xi) , 
\end{equation}
where $\xi = \fourp\cdot\fourx$.
Using equations~(\ref{eq130}) and (\ref{eq131}), we get for equation~(\ref{eq129})
\begin{equation}\label{eq132}
\overline{x_1^i}  \, \overline{s}_0(x) 
= 4\pi \frac{2}{\lambdac^3\Nel} \, x^i \int \nocce(p) \, p^2\,  \rmd p \, \Psi_i(x,\gamma).		
\end{equation}
Analytical expressions for $\Psi_i$ along with asymptotic formulae for different limiting cases can be found in NP94.
For calculating $\dotxc$ and $\Dph(x)$ using equations (\ref{eq113b}) 
we need terms like $\overline{(x_1-x)^i}\, \overline{s}_0(x)$, 
for $i=1,2$, which are simply
\beq \label{eq133}
\overline{(x_1-x)} \, \overline{s}_0(x) 
& = & 4\pi \frac{2}{\lambdac^3\Nel} \, x \int  \nocce(p) \, p^2\, \rmd p \; (\Psi_1 - \Psi_0) ,\\
\label{eq134}
\overline{(x_1-x)^2}\, \overline{s}_0(x)  & = & 4\pi \frac{2}{\lambdac^3\Nel} \,  x^2 \int   \nocce(p) \, p^2\,  \rmd p \; (\Psi_2 - 2\Psi_1 + \Psi_0).	
\eeq

We now rewrite the moments of the electron redistribution function given by equation~(\ref{eq70}) using equation~(\ref{eq15})
\begin{equation}\label{eq135}
\overline{\gamma_1^i} \, \overline{s}_0(p) = \frac{3}{16\pi} \frac{2}{\lambdac^3\Nph} \frac{1}{\gamma}
\int \frac{\rmd ^3 x}{x} \frac{\rmd ^3 x_1}{x_1} \frac{\rmd ^3 p_1}{\gamma_1} \
\gamma_1^i \ \noccx(\vecx) \: F \: \delta^4. 
\end{equation}
We can define quantities analogous to equations~(\ref{eq130}) and (\ref{eq131}):
\begin{equation} 	\label{eq136}
\langle \gamma_1^i \rangle \, s_0(\xi) = \frac{3}{16\pi} \frac{1}{\xi} \int \frac{\rmd ^3 x_1}{x_1} 
\frac{\rmd ^3 p_1}{\gamma_1} \gamma_1^i \: F \: \delta^4 
\end{equation}
and
\begin{equation} \label{eq137}
\Phi_i(x,\gamma) = \frac{1}{4\pi x \gamma^{i+1}} 
\int \rmd^2\omega\ \, \xi \, \langle\gamma_1^i\rangle\,  s_0(\xi) .					
\end{equation}
Equation~(\ref{eq135}) then takes the form
\begin{equation}\label{eq138}
\overline{\gamma_1^i}\, \overline{s}_0(p)  =  4\pi \frac{2}{\lambdac^3\Nph}  \gamma^i \int    \noccx(x) \, 
x^2 \,  \rmd x \ \Phi_i(x,\gamma),				
\end{equation}
while the terms needed for calculating $\dotgammac$ and $\De(\gamma)$ using equation (\ref{Cnum:gd}) are
\beq \label{eq139}
\overline{(\gamma_1-\gamma)}\, \overline{s}_0(p) & = &  4\pi \frac{2}{\lambdac^3\Nph}   \gamma \int    \noccx(x) \, 
x^2 \, \rmd x \; (\Phi_1 - \Phi_0),\\	
\label{eq140}
\overline{(\gamma_1-\gamma)^2} \, \overline{s}_0(p) & =&    4\pi \frac{2}{\lambdac^3\Nph}  \gamma^2 \int   \noccx(x) \,  x^2 \, \rmd x \;
(\Phi_2 - 2\Phi_1 +\Phi_0).												
\eeq

From considerations of energy conservation one would expect a relation between the rates
(\ref{eq133}), (\ref{eq134}) and (\ref{eq139}), (\ref{eq140}).
To see this, consider the quantities $x \, (\Psi_1 - \Psi_0)$ and $\gamma \, (\Phi_1 - \Phi_0)$ entering equations~(\ref{eq133}) and (\ref{eq139}):
\begin{equation} \label{eq141}
x \, (\Psi_1 - \Psi_0) = \frac{1}{4\pi\gamma x} \int \rmd^2\Omega\ \, \xi \, \langle x_1 - x \rangle s_0(\xi) ,	
\end{equation}
where
\begin{equation} \label{eq142}
\langle x_1-x \rangle \, s_0(\xi) = 
\frac{3}{16\pi} \frac{1}{\xi } \int \frac{\rmd ^3 x_1}{x_1} \frac{\rmd ^3 p_1}{\gamma_1} (x_1-x)
\: F \: \delta^4.											
\end{equation}
Analogously for electrons
\begin{equation}\label{eq143}
\gamma \, (\Phi_1 - \Phi_0)  = \frac{1}{4\pi x \gamma} \int \rmd^2\omega\ \,  \xi \,
\langle\gamma_1 - \gamma \rangle \, s_0(\xi) , 
\end{equation}
\begin{equation} \label{eq144}
\langle \gamma_1 - \gamma \rangle\, s_0(\xi)  = 
\frac{3}{16\pi} \frac{1}{\xi } \int \frac{\rmd ^3 x_1}{x_1} \frac{\rmd ^3 p_1}{\gamma_1} (\gamma_1-\gamma)
\: F \: \delta^4.																
\end{equation}
Due to the energy conservation $\delta$-function, $x_1-x = \gamma-\gamma_1$, and we thus get
\begin{equation} \label{eq145}
\langle x_1-x \rangle\, s_0(\xi )  = - \langle \gamma_1 - \gamma \rangle \, s_0(\xi ).
\end{equation}
Also, after performing the integrals over $\rmd ^3 x_1 \, \rmd ^3 p_1$ in equations~(\ref{eq142}) and (\ref{eq144})
the only remaining angle that $\langle x_1 -x \rangle s_0(\xi )$ and $\langle \gamma_1 - \gamma \rangle s_0(\xi)$ can depend on 
is the one between the incoming photon and electron momenta.
Therefore in equations~(\ref{eq141}) and (\ref{eq143}) we can write $\rmd^2\Omega= \rmd^2\omega = 2\pi \rmd\zeta$, $\zeta = \vOmega\cdot\vomega$ and conclude that
\begin{equation} \label{eq146}
x \, (\Psi_1 - \Psi_0) = -\gamma \, (\Phi_1 - \Phi_0).									
\end{equation}
Using the same arguments one can show that
\begin{equation} \label{eq147}
x^2 \, (\Psi_2  - 2 \Psi_1 + \Psi_0) = \gamma^2 \, (\Phi_2 - 2\Phi_1 +\Phi_0).					
\end{equation}
We can thus use the analytic expressions for $\Psi_i$ for calculating the rates
$\dotgammac$ and $\De(\gamma)$ for electrons as well as photons.
Note that since $\Psi_0 = \Phi_0$ we also have analytic expressions for calculating the total scattering cross-section
for electrons through equation~(\ref{eq138}), setting $i = 0$.

\section{Electron-positron pair-production and pair-annihilation rates}
\label{app:pairs}

The quantity  $\Rgg$ entering both the pair-production and the annihilation rate expressions (\ref{eq38_3}) and (\ref{eq38_11}) can be written as
\begin{equation} \label{eq148} 
\Rgg(\gmn,x,x_1) =
\left. \frac{1}{4}\left[ -\sqrt{(x+x_1)^2 - 4\xcm^2} + T(\gmn,x,x_1,\xcm) + T(\gmn,x_1,x,\xcm) \right] \right|_{\xcm^L}^{\xcm^U} .
\end{equation}
The primitive functions are \citep{NL99} 
\beq \label{eq149} 
T(\gmn,x,x_1,\xcm) &=& \frac{\xcm^3}{(xx_1)^{3/2}} \, (xx_1 - 1) \,  \frac{A_0(h) - A(h)}{h}
- \frac{ A(h)}{\xcm \sqrt{x x_1}} 	\nonumber \\
&+& \frac{\xcm}{2 (xx_1)^{3/2}} \left[ \frac{x \, (x_1+x) + \gmn\, (x_1-x) - 2\xcm^2}{A(h)} - 4 x x_1 A_0(h)
\right],									
\eeq
where $h = [(\gmn - x)^2 - 1] \,\xcm^2/xx_1$ and $\xcm$ is the photons' energy in the center of momentum frame.  Upon exchanging $x$ and $x_1$ in the preceding expression one also has to reverse the arguments in function $h$. The quantities $A$ and $A_0$ are identical to the functions defined by equations (\ref{eq128}) and (\ref{eq126}) for Compton scattering. The  expressions similar to (\ref{eq148}) and (\ref{eq149}) have been derived by \citet{Sve82a} and \citet{BS97}. However, their formulae suffer from cancellations when $h$ approaches zero, while in equation (\ref{eq149}) cancellation appears only in the term $[A_0(h) - A(h)]/h$, which can easily be computed via Taylor series for small $h$.

The integration limits in equation (\ref{eq148}) are
\begin{equation} 	\label{eq38_6a}
\xcm^L = \xcm^{-}  \quad \mbox{and} \quad \xcm^U = \min\{ \sqrt{x x_1}, \xcm^{+} \},	
\end{equation}
where
\begin{equation} \label{eq38_6b}
\left( \xcm^{\pm} \right)^2 = \left( \gcm^{\pm} \right)^2 = 
\frac{1}{2} \left[ \gmn \gpl + 1 \pm \sqrt{(\gmn^2 - 1)(\gpl^2 - 1 )} \right] 	
\end{equation} 
and $\gpl=x+x_1-\gmn$ (which follows from the energy conservation).


The lower limit of integration in equation~(\ref{eq38_3}) are expressed as
\begin{equation} \label{eq38_7}
x^{(L)} = \frac{1}{2}\gmn(1 - \bmn) ,  \qquad
x_1^{(L)} = 
\begin{cases} 
x /\{ [2x - \gmn (1 + \bmn)]\gmn (1 + \bmn) \} 
 &\text{if $x > x_+$,} \\
 x /\{ [2x - \gmn (1 - \bmn)]\gmn (1 - \bmn) \} 
&\text{if $x < x_-$,} \\
\gmn + 1 - x &\text{if $x_- \le x \le x_+$,}
\end{cases}
\end{equation}
where we have defined
\begin{equation} 
x_{\pm} = \displaystyle\frac{1}{2}\left[1 + \gmn(1 \pm \bmn)\right].									
\end{equation} 
Because we always have $x_1^{(L)} \ge x^{(L)}$,  the latter sets a lower limit for the energy of either photon for producing
an electron (or positron) of energy $\gmn$.

The lower limits of the momentum integrals in equation~(\ref{eq38_11}) are given as follows \citep{Sve82a}:
\begin{equation}
\gmn^{(L)}  =  
\begin{cases} 
\gamma^{(-)}  &\text{if $x \le 1/2$,} \\
\gamma^{(-)}  &\text{if $1/2 < x < 1$ and $\gpl < \gamma_{\rm B}$,} \\
\gamma^{(+)} &\text{if $x \ge 1$ and $\gpl < \gamma_{\rm B}$,} \\
1                     & \text{in all other cases}, 
\end{cases}
\qquad 
\gpl^{(L)} =   
\begin{cases} 
\gamma_{\rm A}   &\text{if $x < 1/2$,} \\
1  &\text{if $x \ge 1/2$,} \\
\end{cases}
\end{equation}
where
\begin{equation}
\gamma^{(\pm)} = \frac{1}{2} \left( F_{\pm} + \frac{1}{F_{\pm}} \right),	\quad	 F_{\pm} = 2x - \gpl(1 \pm \bpl), \quad
\gamma_{\rm A} = \frac{4 x^2 + 1}{4 x}, \quad \gamma_{\rm B} = \frac{2x^2 - 2x + 1}{2x - 1}.
\end{equation}

The total pair-production cross-section (in units of $\sigmat$)  is given by \citep{GS67,ZDZ88}
\begin{equation}
\sigmapp(x,x_1) = \frac{3}{8} \frac{1}{x^2 x_1^2} \left\{
\frac{2v^2+2v+1}{v+1} \ln{w} - \frac{2(2v+1)\sqrt{v}}{\sqrt{v+1}}
- \ln^2{w}
+ 2 \ln^2{(w+1)}
+ 4\mbox{Li}_2 \left( \frac{1}{w+1} \right)
- \frac{\pi^2}{3}
\right\},
\end{equation}
where
\begin{equation}
v = xx_1 - 1 \quad \mbox{and} \quad w = \frac{\sqrt{v+1} + \sqrt{v}}{\sqrt{v+1} - \sqrt{v}}
\end{equation}
and $\mbox{Li}_2$ is the dilogarithm defined by
\begin{equation}
\mbox{Li}_2 (r) = - \int_0^{r} \frac{\ln(1-s)}{s} \rmd s.
\end{equation}

The total pair annihilation cross-section is found to be \citep[e.g.][]{Sve82a}
\begin{equation}
\sigmapa(\gpl,\gmn) = \frac{3}{8} \, \frac{1}{\gpl \gmn \zpl \zmn} \left. \left[ \bcm^3 \gcm^2 L(\bcm) - 2\gcm^2 + \frac{3}{4} L^2(\bcm)
\right]\right|_{\gcm^{-}}^{\gcm^{+}}.
\end{equation}
Here we have defined $L(\beta) = \ln[(1+\beta)/(1-\beta)]$, while the limits of integration are given by equation~(\ref{eq38_6b}) and $\bcm=\sqrt{1-1/\gcm^2}$.

\section{Cyclo-synchrotron emissivities}
\label{app:synch}

The  cyclo-synchrotron emissivity (here in units s$^{-1}$ str$^{-1}$) at photon energy $x$ in the direction given by angle $\theta$ to the magnetic field for an electron moving at a pitch-angle $\alpha$  with   velocity $\beta=p/\gamma$ is \citep{Pac70}
\begin{equation}  \label{cs01}
\eta (x,\theta,\alpha) = \frac{c}{\lambdac}\alphafs \ x^2 \ \sum_{l=1}^{\infty} 
\left[ \left( \frac{\cos\theta-\beta\cos\alpha}{\sin\theta} \right)^2 J_l^2(z) + 
\beta^2\sin^2\alpha\  {J'_l}^2(z) \right]
\delta \left( l \frac{b}{\gamma} - x [1-\beta\ \cos\alpha\ \cos\theta]\right) ,
\end{equation}
where $\alphafs=e^2/c\hbar$ is the fine-structure constant, 
$b=B/\Bcr$ is magnetic field in units of the critical field $\Bcr= \me^2c^3/(e\hbar)=4.41\times 10^{13}$ G,  
$J_l$ and $J_l'$ are the Bessel function and its derivative, and their argument $z=x p \sin\alpha\ \sin\theta\ /b$.
Averaging over pitch-angle and integrating over $\theta$, we get  the angle-averaged cyclo-synchrotron spectrum
\begin{equation} \label{cs02}
P(x,\gamma) =  \frac{1}{2} \int_{-1}^1 \rmd\cos\alpha \ 2\pi \int_{-1}^1 \rmd\cos\theta \ \eta (x,\theta,\alpha). 
\end{equation}

Direct summation over harmonics works fine for mildly relativistic electrons $\gamma<3$. In this case, we first use the $\delta$-function to integrate over the energy bin, and then  integrate numerically over the angles \citep[see e.g.][]{MM03} and sum over harmonics contributing to a given bin. The same procedure is used for any larger $\gamma$ at photon energies $x$ corresponding to the first 30 harmonics (i.e. $x<30\ b/\gamma$).  At higher $x$, we use two different methods. In the ultra-relativistic regime  $\gamma>10$ we use the angle-averaged relativistic synchrotron spectrum \citep{CS86,GGS88}: 
\begin{equation} \label{eq30}
P(x,\gamma) = \frac{3\sqrt{3}}{\pi} \frac{\sigmat \UB}{\me c}\,  \frac{1}{b} \, \overline{x}^2
\left\{ K_{4/3}(\overline{x}) K_{1/3}(\overline{x}) - \frac{3}{5} \overline{x}
\left[ K^2_{4/3}(\overline{x}) - K^2_{1/3}(\overline{x})
\right] \right\},							
\end{equation}
where $\overline{x} = x/(3\gamma^2 b)$ and $K_y$ is the modified Bessel function.
For   $3<\gamma<10$, we substitute the sum over harmonic in equation (\ref{cs01}) by the integral over $l$ and use 
the $\delta$-function to take it. The   angular integrals are then taken numerically. 
Alternatively we use the approximate formulae proposed by \citet{KGS06}, which ignore harmonics. These give
identical results for the simulations presented in the paper, because low harmonics are self-absorbed. 

All the emissivities $P(x,\gamma)$ are renormalized to guarantee the correct cooling rate given by equation (\ref{eq:gammas}).

\section{Coulomb exchange rates}
\label{app:Coul}

The Fokker-Planck treatment of Coulomb (M{\o}ller) scattering in relativistic plasma was first considered by \citet{DL89}. 
Useful analytical expressions for the energy exchange rate and diffusion coefficient for (small angle) scattering
of test electron of energy $\gamma$ interacting with the background electrons of energy $\gamma_1$ 
have been derived by \citet{NM98}. These expressions (corrected for a few misprints and reorganized) are: 
\begin{equation} 
a(\gamma,\gamma_1) = \frac{3}{4} \frac{c \sigmat \ln{\Lambda}}{\gamma\gamma_1 p p_1} (\gamma_1 - \gamma) \, \chi(\gamma,\gamma_1)
\end{equation}
and
\begin{equation} 
d(\gamma,\gamma_1) = \frac{3}{4} \frac{c \sigmat \ln{\Lambda}}{\gamma\gamma_1 p p_1} 
\Delta(\gamma,\gamma_1)
\end{equation}
where $\ln\Lambda\sim20$ is the Coulomb logarithm,  
\begin{equation} \label{eq:chi_coul}
\chi(\gamma,\gamma_1) = \int_{\gamma_{\rm rel}^{-}}^{\gamma_{\rm rel}^{+}}
\frac{\gammarel^2}{\prel \, (\gammarel-1)} \, \rmd \gammarel = 
\left.\left[ \prel - \frac{\gammarel + 1 }{\prel} + \ln(\gammarel + \prel) 
\right] \right|_{\gamma_{\rm rel}^{-}}^{\gamma_{\rm rel}^{+}}
\end{equation}
and
\beq \label{eq:del_coul}
\Delta(\gamma,\gamma_1) &=& \int_{\gamma_{\rm rel}^{-}}^{\gamma_{\rm rel}^{+}}
\frac{\gammarel^2}{\prel^3}\  (\gamma_{\rm rel}^{+}-\gamma_{\rm rel}) 
(\gamma_{\rm rel}-\gamma_{\rm rel}^{-}) \, \rmd \gammarel \\
&=&
\left. \left\{ - \left(\gamma^2+\gamma_1^2+\frac{1}{2}\right)  \ln(\gammarel + \prel) + 
\frac{1}{\prel} \left[   \gamma_{\rm rel} (\gamma^2+\gamma_1^2)-2 \gamma\gamma_1 \right] + 
\prel  \left(2 \gamma\gamma_1-\frac{\gamma_{\rm rel}}{2}\right)  
\right\}
\right|_{\gamma_{\rm rel}^{-}}^{\gamma_{\rm rel}^{+}}. \nonumber
\eeq
Here $\prel =\sqrt{\gammarel^2-1}$ is the relative momentum and the integration limits are
\begin{equation} 
\gamma_{\rm rel}^{\pm} = \gamma\gamma_1 \pm p p_1.
\end{equation}

We can obtain simple approximate expressions for the energy exchange and diffusion coefficients by approximating the integrals in
equation (\ref{eq:chi_coul}) using one-point trapezoidal and in equation (\ref{eq:del_coul})  a 3-point Simpson's rule:  
\begin{equation}
a(\gamma,\gamma_1) \approx \frac{3}{2} c \sigmat \ln{\Lambda} \, \frac{(\gamma_1 - \gamma) \, \gamma \gamma_1}{(\gamma\gamma_1 - 1)\sqrt{(\gamma\gamma_1)^2 - 1}}
\end{equation}
and
\begin{equation}
d(\gamma,\gamma_1) \approx c \sigmat \ln{\Lambda} \, 
\frac{\gamma\gamma_1 p^2 p_1^2\ \ }{\left[ (\gamma\gamma_1)^2 - 1 \right]^{3/2}} \ , 
\end{equation}
which agree with the exact coefficients reasonably well, except in the region $\gamma \approx \gamma_1$.



\end{document}